\documentclass[aps,print,prb,twocolumn,superscriptaddress]{revtex4-2}

\usepackage[hidelinks]{hyperref}
\hypersetup{
    colorlinks,
    linkcolor={red},
    citecolor={blue},
    urlcolor={blue}
}

\usepackage{balance}
\usepackage{amssymb}
\usepackage{suffix}
\usepackage{mathtools}
\usepackage[utf8]{inputenc}
\usepackage{booktabs}
\usepackage{cases}
\usepackage[multiple]{footmisc}
\usepackage{dcolumn}
\usepackage{color,soul}
\usepackage{rotating}
\usepackage{perpage}
\usepackage{siunitx}
\usepackage{xcolor}
\usepackage{soul}
\usepackage{amsmath}
\usepackage{tikz}
\usepackage[T1]{fontenc}
\usepackage{etoolbox}
\usepackage{graphics}
\usepackage{siunitx}
\usepackage{float}	
\usepackage{collref}
\usepackage{multirow}
\usepackage{mathtools}
\usepackage{bm}
\usepackage{url}

\usepackage{tikz}
\usepackage{tikz-3dplot}
\usepackage{accents}

\makeatletter
\newcommand{\doublewidetilde}[1]{{%
  \mathpalette\double@widetilde{#1}%
}}
\newcommand{\double@widetilde}[2]{%
  \sbox\z@{$\m@th#1\widetilde{#2}$}%
  \ht\z@=.9\ht\z@
  \widetilde{\box\z@}%
}
\makeatother

\usepackage{etoolbox,lipsum}

\DeclarePairedDelimiterX\MeijerM[3]{\lparen}{\rparen}%
{\begin{smallmatrix}#1 \\ #2\end{smallmatrix}\delimsize\vert\,#3}

\newcommand\MeijerG[8][]{%
	\mathbb{G}^{\,#2,#3}_{#4,#5}\MeijerM[#1]{#6}{#7}{#8}}

\WithSuffix\newcommand\MeijerG*[7]{%
	\mathbb{G}^{\,#1,#2}_{#3,#4}\MeijerM*{#5}{#6}{#7}}

\begin{document} 
%Suppressed
 %\title{RKKY interaction in optically driven and gated two-dimensional altermagnets}
 \title{Anisotropic light-tailored RKKY interaction in two-dimensional $d$-wave altermagnets}

\author{Mohsen Yarmohammadi}
\email{mohsen.yarmohammadi@georgetown.edu}
\address{Department of Physics, Georgetown University, Washington DC 20057, USA} 
	\author{Ulrich Z\"{u}licke}
\address{MacDiarmid Institute, School of Chemical and Physical Sciences, Victoria University of Wellington, P.O. Box 600, Wellington 6140, New Zealand}
\address{Dodd-Walls Centre for Photonic and Quantum Technologies, School of Chemical and Physical Sciences, Victoria University of Wellington, P.O. Box 600, Wellington 6140, New Zealand}
    \author{Jamal Berakdar}
\address{Institut f\"ur Physik, Martin-Luther Universit\"at Halle-Wittenberg, 06099 Halle/Saale, Germany}
 \author{Jacob Linder}
	\address{Center for Quantum Spintronics, Department of Physics, Norwegian University of Science and Technology, NO-7491 Trondheim, Norway}
\author{James K. Freericks}
	\address{Department of Physics, Georgetown University, Washington DC 20057, USA}
    
	\date{\today}
	
	\begin{abstract}
     Altermagnets are known in spintronics for their intrinsic spin-splitting and unconventional magnetic responses, particularly to magnetic impurities. However, effectively controlling the magnetic exchange interactions in altermagnets is challenging for practical applications. Here, we propose using circularly polarized light to tune the Ruderman–Kittel–Kasuya–Yosida (RKKY) interaction in two-dimensional $d$-wave altermagnets. Using the real-space retarded Green's functions approach, our results show that while the Heisenberg and Ising exchanges dominate, a notable Dzyaloshinskii–Moriya (DM) interaction also plays a key role. Furthermore, the inherent strength of altermagnetism imprints chirp-like signatures into the magnetic responses, which can be dynamically tuned via light. We mainly demonstrate that gate-induced Rashba spin-orbit coupling is essential in response to light—light selectively and anisotropically adjusts the DM interaction without affecting the other exchanges. Our findings further indicate that rotating the altermagnet by \( 45^\circ \) relative to the light’s polarization direction generates a Dirac-like dispersion and different DM interactions. We finally extract critical thresholds where light reverses DM interactions along one axis or balances both in-plane components. The anisotropic light-driven control of RKKY interactions in 2D altermagnets not only highlights their unique properties but also opens new avenues for engineering tailored magnetic characteristics in spintronic applications. 
	\end{abstract}
	
	\maketitle
	{\allowdisplaybreaks
		
		%\parskip
        %%%%%%%%%%%%%%%%%%%%%%%%%%%%%%%%%%%%%%%%%%%%
		%%%%%%%%%%%%%%%%%%%%%%%%%%%%%%%%%%%%%%%%%%%%
  
		\section{Introduction}
        Altermagnetism introduces a revolutionary paradigm in magnetic materials by merging characteristics traditionally attributed to ferromagnets and antiferromagnets~\cite{PhysRevX.12.040501,PhysRevX.12.040002,PhysRevX.12.031042,doi:10.7566/JPSJ.88.123702,PhysRevX.12.011028,PhysRevB.102.014422,PhysRevB.99.184432,PhysRevMaterials.5.014409,he2025altermagnetismttprimedeltafermihubbardmodel}. This system arises from the interplay of nonrelativistic effects and crystal rotational symmetries, generating momentum-dependent spin splittings, often exhibiting $d$-, $g$-, or even $f$-wave patterns, while the net magnetization cancels out~\cite{doi:10.1073/pnas.2108924118,PhysRevLett.130.216701,PhysRevLett.132.086701,PhysRevLett.132.176702}. This results in band structures with nodal points where time-reversal symmetry is broken despite no net magnetic moment. Experimentally observed in materials like RuO$_2$~\cite{Feng2022,doi:10.1126/sciadv.aaz8809,PhysRevLett.128.197202,weber2024opticalexcitationspinpolarization}~(although altermagnetism in RuO$_2$ is controversial, with no current consensus on its existence in the material) and MnTe~\cite{PhysRevLett.132.036702,PhysRevB.107.L100418,PhysRevB.107.L100417}, and theorized in compounds such as La$_2$CuO$_4$~\cite{PhysRevX.12.031042}, MnF$_2$~\cite{PhysRevB.102.014422}, and Mn$_5$Si$_3$~\cite{PhysRevX.12.040501,reichlová2021macroscopictimereversalsymmetry}, altermagnetism defies traditional classifications, revealing a compensated antiparallel order that sustains robust spin-split electronic states.
        
        The exceptional electronic landscape of altermagnets paves the way for innovative applications in spintronics~\cite{PhysRevLett.130.216701,PhysRevLett.128.197202,PhysRevLett.129.137201,PhysRevLett.134.106801,PhysRevLett.134.106802,PhysRevB.110.235101} and superconducting electronics~\cite{PhysRevLett.131.076003,PhysRevB.108.054511,PhysRevB.108.L060508,PhysRevB.108.184505,PhysRevB.108.224421,Zhang2024}. With anisotropic spin textures, altermagnets enable the generation of spin-polarized currents and give rise to unconventional transport phenomena such as anomalous Hall effects~\cite{doi:10.1126/sciadv.adn0479,Feng2022,doi:10.1126/sciadv.aaz8809}, giant and tunneling magnetoresistance~\cite{PhysRevX.12.011028}, and charge-spin conversion~\cite{Bose2022}. These properties challenge conventional understandings of magnetic order and offer a versatile platform for both fundamental research and technological innovation.
        
		In altermagnets, where unconventional spin-splitting arises despite the absence of a net magnetization, magnetic impurities interact via the anisotropic and oscillatory Ruderman–Kittel–Kasuya–Yosida (RKKY) mechanism~\cite{10.1143/PTP.16.45,*PhysRev.106.893,*PhysRev.96.99,lee2024magneticimpuritiesaltermagneticmetal,PhysRevB.110.054427}. Additionally, the application of external magnetic fields or the introduction of Rashba spin-orbit coupling, achievable through substrate engineering or interface modifications, offers an avenue to control the local spin texture precisely and, consequently, the magnetic coupling. Such control over impurity interactions could lead to the advancement of spintronic devices, where the manipulation of localized magnetic moments enables the realization of novel functionalities in memory storage, logic circuits, and quantum information processing. While some studies have explored the RKKY interaction in altermagnets~\cite{lee2024magneticimpuritiesaltermagneticmetal,PhysRevB.110.054427}, propositions to control this interaction remain unexplored. 

While static approaches like Zeeman field~\cite{PhysRevB.110.054427} have their merits, dynamic control, often through Floquet engineering, offers a versatile way to modify the electronic band structure of altermagnets by inducing transient changes in momentum-dependent spin splitting. Unlike traditional methods, light provides a noninvasive, ultrafast, and spatially precise approach to control the RKKY interaction. By tuning the frequency, intensity, or polarization, one can precisely adjust the strength and oscillatory behavior of the RKKY interaction~\cite{PhysRevB.111.014440,PhysRevB.107.054439,PhysRevResearch.2.033228,PhysRevB.110.035307}. 

Given this background, this work addresses a crucial question: How can \textit{circularly polarized light} modulate the unconventional phenomena of altermagnetism, taking crystallographic orientations into account, and in turn, influence the RKKY interaction between two magnetic impurities embedded within the material? Specifically, we focus on a two-dimensional~(2D) altermagnet. We also incorporate gate voltages applied to both the top and bottom of the layer, thereby bringing the system closer to practical realizations, as shown in Fig.~\ref{f1}. Experimentally, one would apply circularly polarized light in tandem with gate electrodes, thereby selectively modulating the RKKY interaction. Although achieving precise control over light, frequency, and gate voltage presents technical challenges, recent advances in ultrafast optics and nanoscale device fabrication suggest that such experiments are within reach~\cite{Higuchi2017,doi:10.1126/science.1239834,McIver2020,PhysRevResearch.2.043408,liu2024evidencefloquetelectronicsteady}. 

We choose to focus on \( d \)-wave altermagnets because their symmetry produces a well‑defined sign-changing spin splitting with fewer nodal points than the more intricate \( g \)- and \( f \)-wave states, resulting in robust spin polarization. Moreover, \( d \)-wave states can generate sizeable spin splittings without net magnetization, making them ideal for spintronic applications that require the manipulation of spin-dependent transport without stray magnetic fields. Additionally, as a lower‑order symmetry, \( d \)-wave is easier to stabilize in 2D systems and is less sensitive to impurities, while still offering sufficient anisotropy for novel spin phenomena and maintaining a manageable electronic structure for both theoretical and experimental studies.

The structure of this paper is as follows. In Sec.~\ref{s2}, we introduce the model Hamiltonian, incorporating both the gate and light effects. Section~\ref{s3} details the calculation of the RKKY interaction. In Sec.~\ref{s4}, we present and discuss the results of our analysis. Finally, we provide a summary in Sec.~\ref{s5}.

        %%%%%%%%%%%%%%%%%%%%%%%%%%%%%%%%%%%%%%%%%%%%
		%%%%%%%%%%%%%%%%%%%%%%%%%%%%%%%%%%%%%%%%%%%%
		\begin{figure}[t]
			\centering
		\includegraphics[width=0.9\linewidth]{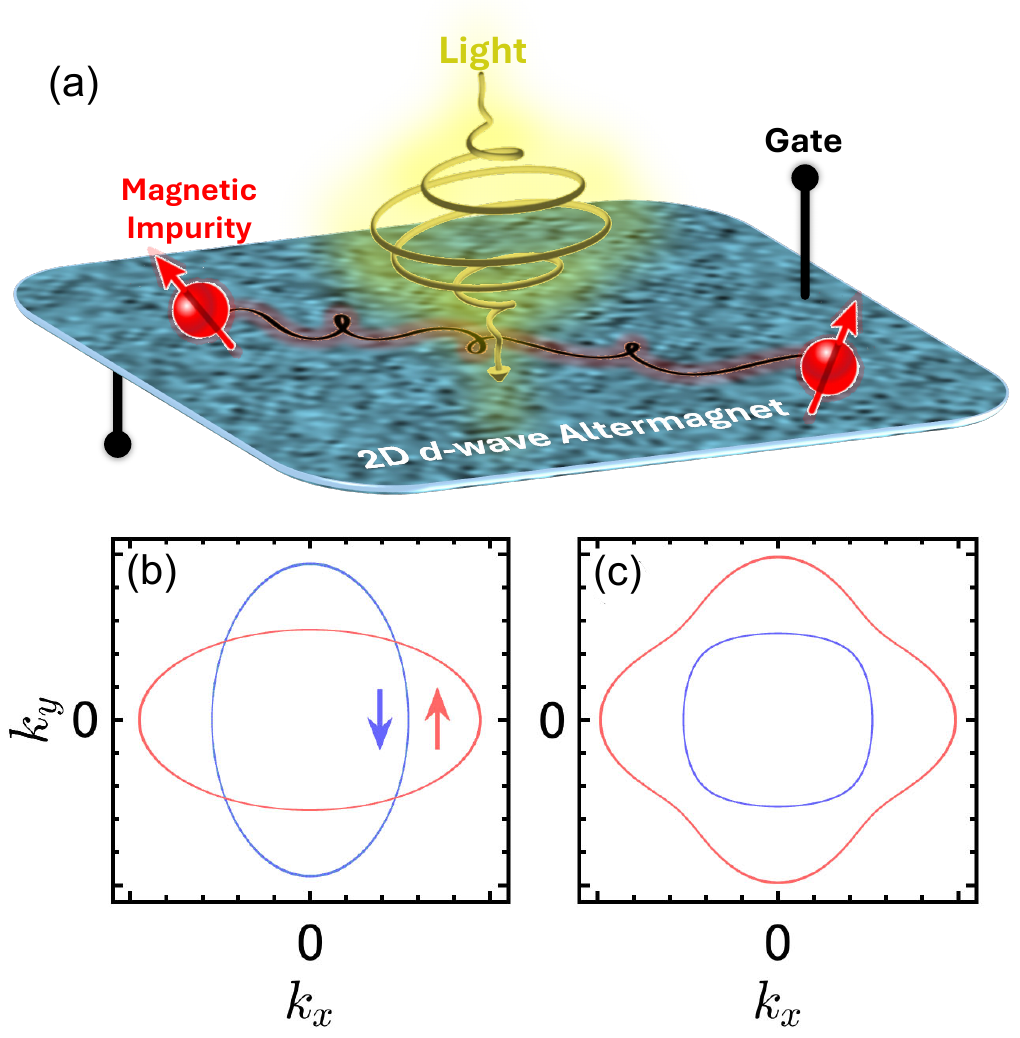}
			\caption{(a) Schematic depiction of a gated (generating Rashba spin-orbit coupling) 2D $d$-wave altermagnet under the influence of circularly polarized light, with two magnetic impurities doped onto the surface to induce the RKKY interaction. The altermagnet is assumed to be an interface between two gate electrodes (which are not shown here to avoid adding complexity to the figure). The spin-splitting effect is shown in the constant energy contour of (b) pristine~(in the absence of driving and gating) and (c) gated ($\lambda = 0.5$ eV$\cdot$\AA) 2D $d$-wave altermagnets with altermagnetism strength of $\beta = 0.5$. Gating can change the electron density, directly affecting the interaction between an electron’s spin and its orbital motion.} 
			\label{f1}
		\end{figure} 
		\section{Model Hamiltonian}\label{s2}
        We characterize the electronic properties of the 2D $d$-wave altermagnet, as illustrated in Fig.~\ref{f1}(a), by a generic two-band model Hamiltonian, which is expressed as $\mathcal{H} = \sum_{\vec{k}} \mathcal{H}_{\vec k}$, where
\begin{align}
    \mathcal{H}_{\vec k} = \alpha_{\vec{k}}\, \sigma_0 + \beta_{\vec{k}}\, \sigma_z\,,
\end{align}with 
\begin{align}\label{eq_2}
    \alpha_{\vec{k}} = \frac{\hbar^2 k^2}{2\,m}\qquad{\rm and}\qquad \beta_{\vec{k}} = \frac{\hbar^2 \beta (k_x^2 - k_y^2)}{2\,m}\, ,
\end{align}where $m$ represents the electron mass and $0 < \beta < 1$ quantifies the dimensionless strength of the altermagnetic interaction. It is important to highlight that $\beta_{\vec k}$ primarily governs the intrinsic spin splitting in altermagnets and the anisotropy of the Fermi surfaces for each spin. Here, $k^2 = |\vec{k}|^2 = k_x^2 + k_y^2$ denotes the magnitude of the crystal momentum in two dimensions. Furthermore, $\sigma_0$ is the $2\times2$ identity matrix, and $\sigma_\ell $ for $\ell \in \{x, y, z\}$ are the Pauli matrices, representing the band-electron spin. Diagonalizing this Hamiltonian results in the electronic band structure, revealing intrinsic spin-splitting features at all energy levels. The matrix is already diagonal, with eigenvalues of $\alpha_{\vec k}+\beta_{\vec k}$ for spin-up and $\alpha_{\vec k}-\beta_{\vec k}$ for spin-down.  At a given energy level, the top view of the dispersion for spin-up and spin-down states produces Fig.~\ref{f1}(b), clearly showing the significant splitting between the dispersions of the spin states. Each spin state is individually governed by a \(C_2\) rotational symmetry, while there exists a \(C_4\) rotational symmetry between the spin states. This means that a 90$^\circ$ rotation in reciprocal space for a given spin leads to the other spin state.

The applied gate voltage breaks the out-of-plane inversion symmetry, a crucial factor in generating Rashba spin-orbit coupling (RSOC) through relativistic effects proportional to $(\vec{\nabla} V \times \vec k) \cdot \vec \sigma$, where the derivative of potential $V$ is related to the effective electric field along the $z$-axis~\cite{winkler2003spin}. Beyond enabling precise control of carrier density, the gate voltage also fine-tunes the spin splitting driven by Rashba effects. Such RSOC generation leads to the following Hamiltonian \begin{align}
    \mathcal{H}^{\rm R}_{\vec k} = \alpha_{\vec{k}}\, \sigma_0 + \beta_{\vec{k}}\, \sigma_z + \lambda\, (k_x\,\sigma_y - k_y\,\sigma_x)\,,
\end{align}where $\lambda$ indicates the strength of the gate-induced RSOC. This term couples the electron's spin to its momentum, breaking inversion symmetry in two dimensions and resulting in modulation of the inherent spin-splitting effect in the band structure, as shown in Fig.~\ref{f1}(c) with $\lambda = 0.5$ eV$\cdot$\AA. Once the gate is turned on,  \(C_2\) symmetry maintains for each spin, and the inherent \(C_4\) symmetry between the spin states is broken.

  The RSOC induced by gate voltages is expected to be weak in altermagnets because many altermagnets, such as Mn\(_5\)Si\(_3\)~\cite{PhysRevX.12.040501,reichlová2021macroscopictimereversalsymmetry} and RuO\(_2\)~\cite{Feng2022,doi:10.1126/sciadv.aaz8809,PhysRevLett.128.197202}, are primarily composed of \(3d\) transition metals, which exhibit weaker atomic SOC compared to heavier elements. Additionally, unlike strong Rashba systems that rely on large structural inversion asymmetry, many altermagnets retain bulk inversion symmetry or only weakly break it. Another reason would be the exchange-driven (altermagnetic) spin splitting which is usually much larger than the relativistic corrections leading to RSOC. These factors collectively suggest that RSOC ($\lambda$) induced by gate voltages in altermagnets is significantly weaker than in conventional Rashba materials.

 As detailed in Appendix~\ref{ap_0l}, the linearly polarized light only causes an energy shift in the system—a point we do not further explore. Appendix~\ref{ap_0e} also demonstrates that the modulation in presence of an elliptically polarized light remains as circularly polarized light, with only the scale of the modulations being altered and a negligible additional mass term due to the difference in the amplitudes of the \( x \) and \( y \) components in the vector potential. Therefore, we proceed with the circularly polarized light. The right circularly polarized light~(see Appendix~\ref{ap_0} for left polarization) is modeled as a time-dependent vector potential, given by $\vec{A}(t) = A_0\, [\sin(\Omega_{\rm d}\, t),\cos(\Omega_{\rm d}\, t)]$, where $A_0 = \mathcal{E}_0/\sqrt{2}\, \Omega_{\rm d}$, with $\mathcal{E}_0$ and $\Omega_{\rm d}$ representing the amplitude and frequency of the light, respectively, see Fig.~\ref{f1}(a). This vector potential couples to the Hamiltonian through the minimal coupling approach ($\vec{k} \to \vec{k} - e\,\vec{A}(t)$, where $e$ is the electron's charge), inducing transitions between the system's eigenstates. Given the time-periodic nature of the Hamiltonian, $\mathcal{H}_{\vec k}(t) = \mathcal{H}_{\vec k}(t + T)$, where $T = 2 \pi/\Omega_{\rm d}$ is the period, we employ the Floquet-Bloch theorem~\cite{GRIFONI1998229,PLATERO20041} to address such time-dependent systems. Within this framework, the Floquet Hamiltonian is calculated and diagonalized to unveil the effects of the light-induced perturbation. 

When the light amplitude is weaker than the light frequency, we can perform a high-frequency expansion to simplify the Floquet Hamiltonian. We set the light frequency to a fixed value of \( \Omega_{\rm d} = 0.35 \) eV, which is larger than the low-energy bandwidth relevant for electronic excitations discussed throughout the paper. In this regime, the Floquet sidebands are sufficiently well-separated, preventing band crossings that would otherwise lead to gap openings. As a result, interband electron transitions are suppressed, and the hybridization between the Floquet bands is diminished. To derive the effective Floquet Hamiltonian in this limit, we employ a perturbative technique known as the van Vleck inverse-frequency expansion, which leads to the following expression~\cite{PhysRevB.84.235108,PhysRevB.82.235114,PhysRevB.84.235108,PhysRevLett.110.026603}: \begin{align}
  \mathcal{H}^{\rm eff}_{\vec{k}} \simeq {} \mathcal{H}^{\rm F}_0 +\frac{\left[\mathcal{H}^{\rm F}_{-1},\mathcal{H}^{\rm F}_{+1}\right]}{\Omega_{\rm d}}\, ,
\end{align}where $\mathcal{H}^{\rm F}_0 = {} \frac{1}{T} \int^T_0 \mathcal{H}^{\rm R}_{\vec k- e \vec{A}}(t)\, dt$ and $\mathcal{H}^{\rm F}_{\pm 1} = {} \frac{1}{T} \int^T_0 \mathcal{H}^{\rm R}_{\vec k- e \vec{A}}(t)\, e^{\pm i \,\Omega_{\rm d} t} dt$. The van Vleck expansion is designed to capture the slow (or averaged) dynamics by providing a time-independent effective Hamiltonian. This effective Hamiltonian governs the long-time evolution, particularly when observed stroboscopically (i.e., at multiples of the driving period) and fast oscillations are neglected.

By evaluating these integrals~(see Appendix~\ref{ap_0} for details) and defining the parameters $\mathcal{A}_0 = e A_0 \lambda$ and $\Delta = \mathcal{A}_0^2/2\,\Omega_{\rm d}$, we ultimately find the effective Floquet Hamiltonian:\begin{align}\label{eq_5}
  & \mathcal{H}^{\rm eff}_{\vec{k}} \simeq {} \left(\alpha_{\vec{k}}+\frac{\hbar^2\,\Delta\,\Omega_{\rm d}}{m\,\lambda^2}\right)\, \sigma_0 + \left(\beta_{\vec{k}}+2\Delta\right)\, \sigma_z \notag \\ {} &+ \left[\left(\lambda- \frac{2\hbar^2\,\Delta \,\beta}{m\,\lambda}\right)\, k_x\,\sigma_y - \left(\lambda+ \frac{2\hbar^2\,\Delta \,\beta}{m\,\lambda}\right)k_y\,\sigma_x \right] .
    \end{align}The quantities $\alpha_{\vec k}$ and $\beta_{\vec k}$ are already defined in Eq.~\eqref{eq_2}. Therefore, light-induced effects occur in the modulation of all terms. It should be emphasized that the term \(2\Delta \sigma_z\) represents the light-induced magnetization, resulting from the inverse Faraday effect~\cite{PhysRevLett.15.190,PhysRevB.110.094302} and solely from the RSOC term. However, when $\Delta$ is considered alone, such as in \( 2\Delta \sigma_z \) term, it becomes negligible due to the smallness of \( \lambda^2 \). However, when taken together with the altermagnetic strength \( \beta \) and the bare RSOC strength \( \lambda \), such as in the ratio \( 2\,\Delta \beta / \lambda \) of the third term, it is not negligible.\begin{figure}[t]
    \centering
    \includegraphics[width=0.8\linewidth]{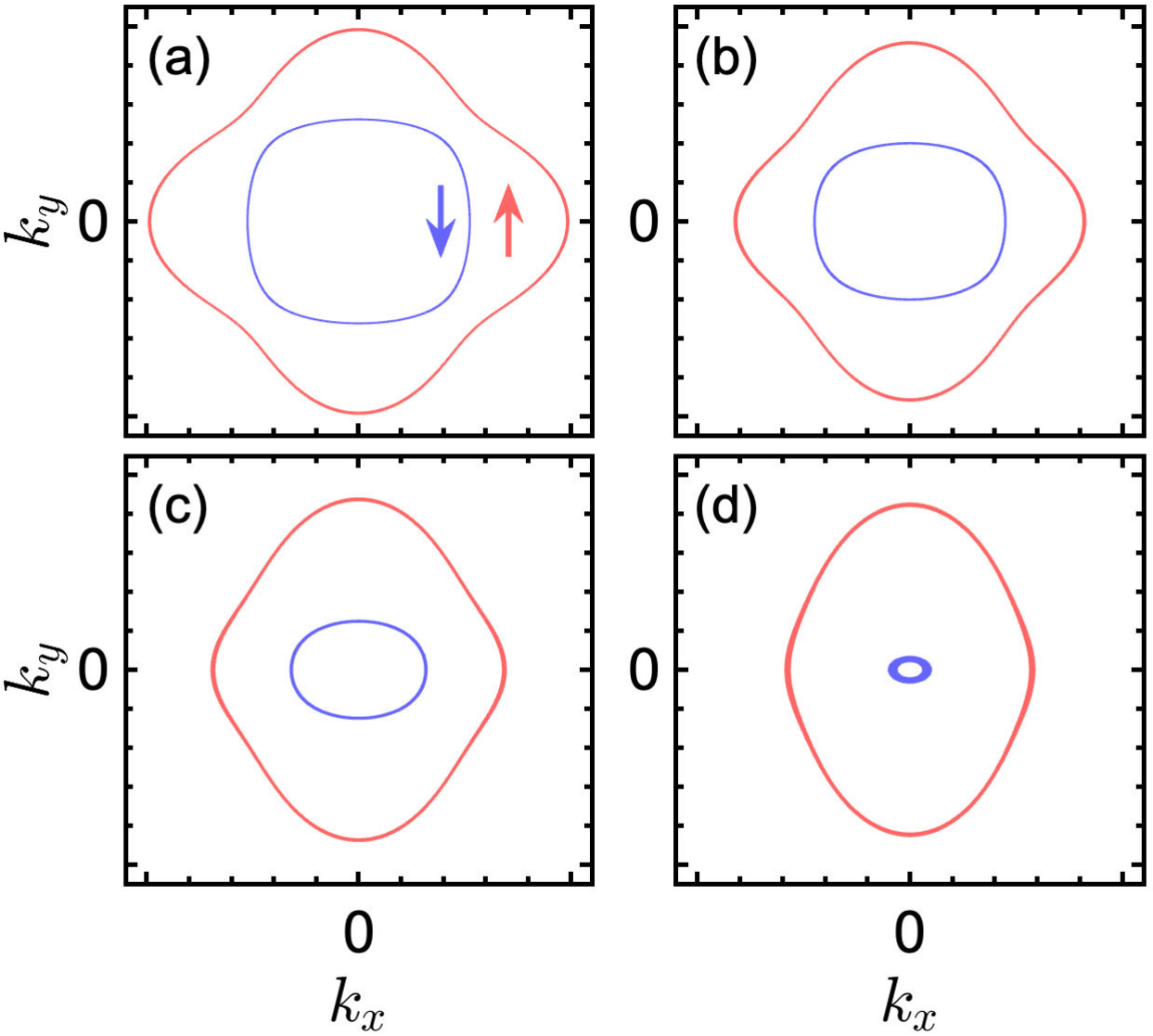}
    \caption{The effect of circularly polarized light~(averaged over the Floquet period via the van Vleck high-frequency expansion) on the spin-splitting through the constant energy contour of a gated ($\lambda = 0.5$ eV$\cdot$\AA) 2D $d$-wave altermagnet with $\beta = 0.5$ at (a) $\Delta = 0$, (b) $\Delta = 0.2$ eV, (c) $\Delta = 0.4$ eV, and (d) $\Delta = 0.6$ eV. We fix the light frequency at $\Omega_{\rm d} = 2$ eV and set $\hbar = m = 1$. The gate-induced \(C_4\) symmetry for each spin is broken by the light, resulting in the generation of a \(C_2\) symmetry similar to the pristine phase, but with a significantly light-dressed Fermi surface.} 
    \label{f3n}
    \end{figure} 
    
    While the diagonal terms are consistent with those reported in previous studies~\cite{PhysRevB.79.081406,doi:10.1063/1.3597412,PhysRevB.83.245436,PhysRevB.85.155449,PhysRevResearch.2.033228,PhysRevB.107.054439,PhysRevB.110.035307}, the off-diagonal terms, which play a crucial role in tuning the anisotropic RSOC have not been explored before. In contrast to 2D Rashba systems without altermagentic features, this modulation essentially requires the altermagnetism term $\beta_{\vec k} \sigma_z$. The observation that altermagnetism, rather than merely time-reversal symmetry breaking (as in a simple Zeeman model like \(\beta \sigma_z\)), resulting in an enhancement of the inversion-symmetry breaking, is required to achieve an anisotropic modulation of the RSOC is a key insight in this study. Thus, the spinor structure---and consequently the orbital weight, due to spin-orbit entanglement---changes with momentum. This modulation effectively rearranges the orbital texture in momentum space~\cite{aase2025orbitalsplittereffectspatial}. For left circularly polarized light, only the sign of the light exchange energy changes in the second and third terms of Eq.~\eqref{eq_5}, i.e., \(\Delta \to -\Delta\), while the modulation remains anisotropic for RSOCs. See Appendix~\ref{ap_0} for details.

Similarly, by diagonalizing the light-induced Hamiltonian, in Fig.~\ref{f3n}, we observe how the light modulates the electronic band structure and the propagation of waves for spin-up and spin-down states. This results in a clear alteration of the Fermi surface, which in turn influences the RKKY interaction. Turning on the light breaks the \(C_4\) symmetry between spins, while it still maintain the \(C_2\) symmetry for each spin.

This result can be most fundamentally understood from the interplay of the multipolar orders underlying the $d$-wave altermagnet (a magnetic octupole density~\cite{PhysRevX.14.011019,PhysRevB.107.155201}) and light (a magnetotoroidal dipole~\cite{tav61}). The new compound-multipole order arising from the combination of these two is an electrotroroidal quadrupole (odd under spatial inversion, even under time inversion) whose $xy$ components renormalize the dipolar Rashba term. Generally, the interaction of Floquet light with a crystalline multipole density will result in the emergence of a new multipole order with rank reduced by one and opposite behavior under space and time inversion. See also Refs.~\cite{Hayami2024,Hayami2025} for a related discussion.
    
    \section{RKKY interaction}\label{s3}
In Fig.~\ref{f1}(a), we place two magnetic impurities, $\vec{S}_1$ and $\vec{S}_2$, at sites $\vec{R}_1$ and $\vec{R}_2$ with a separation vector $\vec{R} = \vec{R}_2 - \vec{R}_1$ within the layer. The indirect cross-talk between them is mediated by the itinerant electron's spins in the host system, as elegantly captured by the RKKY theory~\cite{10.1143/PTP.16.45,*PhysRev.106.893,*PhysRev.96.99}. In the present context, these itinerant electrons are represented by the optically driven and gate-controlled carriers in a 2D $d$-wave altermagnet, as described by the expression in Eq.~\eqref{eq_5}. By applying second-order perturbation theory, the RKKY Hamiltonian can be expressed as\begin{align}\label{eq_6}
			\mathcal{H}_{\rm RKKY} = -\frac{ J^2}{\pi} {\rm Im} \int^0_{-\infty} d \,\mathcal{E} \,\chi(\vec{R},\mathcal{E})\, ,
		\end{align}where $J$ is the bare magnetic exchange coupling between localized moments and itinerant-electron spins, and the spin susceptibility is given by
        \begin{align}\label{eq_7}
            \chi(\vec{R},\mathcal{E}) = {} {\rm Tr}\,[(\vec{S}_1\cdot \vec{\sigma})\,G(\vec{R},\mathcal{E})(\vec{S}_2\cdot \vec{\sigma})\,G(-\vec{R},\mathcal{E})]\, ,
        \end{align}with $G$ representing the retarded Green's functions. 

        To obtain the real-space retarded Green's functions, we first require the momentum-space Green's functions, which are expressed as $G(\vec{k}, \mathcal{E}) = ([\mathcal{E} + i\eta]\sigma_0 - \mathcal{H}^{\rm eff}_{\vec{k}})^{-1}$, where $\eta$ is an infinitesimally small parameter. After straightforward calculations and setting $\lambda^2 \to 0$ where necessary, we find\begin{align}
G(\vec{k}, \mathcal{E}) \simeq {} \begin{pmatrix}
            \frac{1}{\mathcal{E} + i\eta -\widetilde{\alpha}_{\vec k} - \beta_{\vec k}} && \frac{-\lambda_y\,k_y - i \lambda_x\,k_x}{[\mathcal{E} + i\eta -\widetilde{\alpha}_{\vec k}]^2 - \beta^2_{\vec k}}\\\\
             \frac{-\lambda_y\,k_y + i \lambda_x\,k_x}{[\mathcal{E} + i\eta-\widetilde{\alpha}_{\vec k}]^2 - \beta^2_{\vec k}} && \frac{1}{\mathcal{E} + i\eta-\widetilde{\alpha}_{\vec k} + \beta_{\vec k}}
            \end{pmatrix}\, ,
        \end{align}where $\widetilde{\alpha}_{\vec k} = \alpha_{\vec k} +\frac{\hbar^2\,\Delta\,\Omega_{\rm d}}{m\,\lambda^2}$ and\begin{align}
                \lambda_x = \lambda - \frac{2\hbar^2\,\Delta \,\beta}{m\,\lambda}\quad{\rm and}\quad\lambda_y = -\lambda-\frac{2\hbar^2\,\Delta \,\beta}{m\,\lambda}. 
            \end{align}This expression can be succinctly represented as\begin{align}\label{eq_9}
            G(\vec{k}, \mathcal{E}) = {} G_0(\vec{k}, \mathcal{E})\,\sigma_0 + \vec{G}(\vec{k}, \mathcal{E})\cdot \vec{\sigma}\, ,
        \end{align}with\begin{subequations}
            \begin{align}
                G_0(\vec{k}, \mathcal{E}) = {} & \frac{\mathcal{E} + i\eta -\widetilde{\alpha}_{\vec k}}{[\mathcal{E} + i\eta-\widetilde{\alpha}_{\vec k}]^2 - \beta^2_{\vec k}}\, ,\\
                G_x(\vec{k}, \mathcal{E}) = {} &\frac{\lambda_y\,k_y}{[\mathcal{E} + i\eta-\widetilde{\alpha}_{\vec k}]^2 - \beta^2_{\vec k}}\, ,\\
                G_y(\vec{k}, \mathcal{E}) = {} &\frac{\lambda_x\,k_x}{[\mathcal{E} + i\eta-\widetilde{\alpha}_{\vec k}]^2 - \beta^2_{\vec k}}\, ,\\
                G_z(\vec{k}, \mathcal{E}) = {} & \frac{\beta_{\vec k}}{[\mathcal{E} + i\eta-\widetilde{\alpha}_{\vec k}]^2 - \beta^2_{\vec k}}\, .
            \end{align}
        \end{subequations}The real-space version of the Green's functions can be easily obtained by performing the Fourier transform $G_n(\vec{R}, \mathcal{E}) = \frac{1}{4\,\pi^2} \int d^2 k\,G_n(\vec{k}, \mathcal{E}) \,e^{i\,\vec{k}\cdot\vec{R}}$, where $n \in\{0,x,y,z\}$. Accordingly, we obtain the required expression for $ G(\vec{R}, \mathcal{E})$ as needed in Eq.~\eqref{eq_7}. Finally, the RKKY Hamiltonian in Eq.~\eqref{eq_6} is given by{\small\begin{align}\label{eq_12}
            \mathcal{H}_{\rm RKKY} = \sum_{\ell,m} \mathcal{J}_{\ell m}(R) [S_1^\ell S_2^m\hspace*{-0.05cm}+\hspace*{-0.05cm}S_1^m S_2^\ell] \hspace*{-0.05cm}+\hspace*{-0.05cm} \vec{\mathcal{J}}_{{\rm DM}}(R) \cdot (\vec{S}_1 \times \vec{S}_2).
        \end{align}}These terms form isotropic/symmetric in-plane Heisenberg interactions, ({$\mathcal{J}_{xx} = \mathcal{J}_{yy}$), Ising interaction ($\mathcal{J}_{zz}$), anisotropic interactions~($\mathcal{J}_{yz}$, $\mathcal{J}_{xz}$), and anisotropic/antisymmetric Dzyaloshinskii–Moriya~(DM) interactions~($\mathcal{J}_{{\rm DM},x}$, $\mathcal{J}_{{\rm DM},y}$). These interactions give rise to unique spin configurations that could play a role in magnetic memory devices and impact spin-wave dynamics, potentially enhancing signal processing technologies~\cite{CAMLEY2023100605}. 
        
        The exchange interactions are given by:\begin{widetext}\begin{subequations}\label{eq_13}
            \begin{align}
                \mathcal{J}_{xx}(R) = & -\frac{ J^2}{\pi} {\rm Im} \int^0_{-\infty} d \,\mathcal{E} \,[G_0^2(R,\mathcal{E}) + G_x(R,\mathcal{E})G_x(-R,\mathcal{E}) - G_y(R,\mathcal{E})G_y(-R,\mathcal{E}) - G_z(R,\mathcal{E})G_z(-R,\mathcal{E})]\, ,\\
                \mathcal{J}_{yy}(R) = & -\frac{ J^2}{\pi} {\rm Im} \int^0_{-\infty} d \,\mathcal{E} \,[G_0^2(R,\mathcal{E}) - G_x(R,\mathcal{E})G_x(-R,\mathcal{E}) + G_y(R,\mathcal{E})G_y(-R,\mathcal{E}) - G_z(R,\mathcal{E})G_z(-R,\mathcal{E})]\, ,\\
                \mathcal{J}_{zz}(R) = &-\frac{ J^2}{\pi} {\rm Im} \int^0_{-\infty} d \,\mathcal{E} \,[ G_0^2(R,\mathcal{E}) - G_x(R,\mathcal{E})G_x(-R,\mathcal{E}) - G_y(R,\mathcal{E})G_y(-R,\mathcal{E}) + G_z(R,\mathcal{E})G_z(-R,\mathcal{E})]\, ,\\
                \mathcal{J}_{xy}(R) = &-\frac{ 2\,J^2}{\pi} {\rm Im} \int^0_{-\infty} d \,\mathcal{E} \,[G_x(R,\mathcal{E})G_y(-R,\mathcal{E}) + G_x(-R,\mathcal{E})G_y(R,\mathcal{E})]\,,\\
                \mathcal{J}_{yz}(R) = &-\frac{ 2\,J^2}{\pi} {\rm Im} \int^0_{-\infty} d \,\mathcal{E} \,[G_y(R,\mathcal{E})G_z(-R,\mathcal{E}) + G_y(-R,\mathcal{E})G_z(R,\mathcal{E})]\,,\\
                \mathcal{J}_{xz}(R) = &-\frac{ 2\,J^2}{\pi} {\rm Im} \int^0_{-\infty} d \,\mathcal{E} \,[G_x(R,\mathcal{E})G_z(-R,\mathcal{E}) + G_x(-R,\mathcal{E})G_z(R,\mathcal{E})]\,,\\
                \mathcal{J}_{{\rm DM},x}(R) = &-\frac{J^2}{\pi} {\rm Im} \int^0_{-\infty} d \,\mathcal{E} i\, G_0(R,\mathcal{E})[G_x(-R,\mathcal{E})-G_x(R,\mathcal{E})]\,,\\
                \mathcal{J}_{{\rm DM},y}(R) = &-\frac{J^2}{\pi} {\rm Im} \int^0_{-\infty} d \,\mathcal{E} i \,G_0(R,\mathcal{E})[G_y(-R,\mathcal{E})-G_y(R,\mathcal{E})]\,,\\
                \mathcal{J}_{{\rm DM},z}(R) = &-\frac{J^2}{\pi} {\rm Im} \int^0_{-\infty} d \,\mathcal{E} i \,G_0(R,\mathcal{E})[G_z(-R,\mathcal{E})-G_z(R,\mathcal{E})]\,.
            \end{align}
        \end{subequations}\end{widetext}
        
        It should be noted that since $G_x(R,\mathcal{E})$ and $G_y(R,\mathcal{E})$ are proportional to $\lambda$, setting $\lambda^2 \approx 0$ leaves only the $G_0(R,\mathcal{E})$ and $G_z(R,\mathcal{E})$ contributions with $\mathcal{J}_{xx}(R) = \mathcal{J}_{yy}(R) \neq \mathcal{J}_{zz}(R)$ and $\mathcal{J}_{xy}(R) = 0$. The closed-form analytical expressions of the real-space Green's functions are given by (see Appendix~\ref{ap_1nn} for derivations):\begin{subequations}\label{eq_14}
    \begin{align}
        G_0(R,\mathcal{E}) = {} &\frac{-m}{4\hbar^2\,\sqrt{1-\beta^2}}
\left[H_0^{(1)}(k_+R)+H_0^{(1)}(k_-R)\right]\, ,\\
G_x(R,\mathcal{E})= {} &\frac{-i\,\lambda_y\, m\,e^{i\phi}}{4\hbar^4\sqrt{1-\beta^2}}\,
\left[\frac{H_1^{(1)}(k_+R)}{k_+}-\frac{H_1^{(1)}(k_-R)}{k_-}\right]\, ,\\
G_y(R,\mathcal{E})= {} &\frac{-\lambda_x\, m\,e^{i\phi}}{4\hbar^4\sqrt{1-\beta^2}}\,
\left[\frac{H_1^{(1)}(k_+R)}{k_+}+\frac{H_1^{(1)}(k_-R)}{k_-}\right]\, ,\\
G_z(R,\mathcal{E}) = {} &\frac{-m\,\beta\,\cos(2\phi)}{4\hbar^2\sqrt{1-\beta^2}}\,
\left[\frac{H_2^{(1)}(k_+R)}{k_+^2}+\frac{H_2^{(1)}(k_-R)}{k_-^2}\right] ,
\end{align}
\end{subequations} where\begin{align}
    k_\pm = {}\sqrt{\frac{2m\left(\mathcal{E}-\frac{\hbar^2\,\Delta\,\Omega_{\rm d}}{m\,\lambda^2}\right)}{\hbar^2(1\pm \beta)}}\,,
\end{align}and $\phi$ represents the direction of the separation vector between the impurities in the $xy$-plane. Note that, by symmetry, we achieve $\mathcal{J}_{{\rm DM},z}(R) = {} 0$ because $H^{(1)}_2(-z) = H^{(1)}_2(z)$ as well as $\mathcal{J}_{yz}(R) = \mathcal{J}_{xz}(R) = 0$ because $H^{(1)}_1(-z) = -H^{(1)}_1(z)$. Thus, only four terms remain, including the Heisenberg, Ising, and in-plane DM terms. The altermagnetism strength ($\beta$) and the light ($\Delta$) serve as crucial tunable parameters that modulate the RKKY responses, pivotal in the control and manipulation of the system's magnetism.

We note that light-induced effects are sensitive to the crystallographic orientation of the altermagnet~\cite{PhysRevB.108.054511}. For instance, the corresponding light-induced effective Hamiltonian and RKKY interaction of a 45$^\circ$ rotation of the illuminated surface is given in Appendix~\ref{ap_00}. For such a special rotation, RSOC term is invariant, while the altermagnetism term is rotated from \( \beta_{\vec{k}} = \frac{\hbar^2\beta}{2\,m}(k_x^2 - k_y^2)\,\sigma_z \) to \( \theta_{\vec{k}} = \frac{\hbar^2 \beta}{m} k_x k_y\,\sigma_z \), which in turn, turns on a Dirac-like dispersion, $\vec{k} \cdot \vec{\sigma}$, stemming from the light. We do not explore additional rotations here, as they can be easily obtained by rotating both real- and spin-space in the same manner.
    
 \section{Results and discussion}\label{s4}Although numerical results could be used to cover all regimes of \( R \) with the Hankel functions themselves, we choose to proceed with their asymptotic forms in order to obtain analytical expressions that can be easily compared with previous studies. However, we focus on large impurity separations ($k R \gg 1$) in this work for the following reasons: (i) The standard derivation of the RKKY interaction relies on second-order perturbation theory, which remains well-controlled at large impurity separations due to the weak coupling, ensuring that higher-order corrections are negligible. In contrast, for closely spaced impurities, the perturbative expansion may break down, necessitating the inclusion of additional effects such as direct exchange interactions or local modifications of the band structure. (ii) When studying universal features that depend only on the low-energy electronic structure of 2D materials like our case, the long-range regime is the natural focus. As a consequence, we use $ H^{(1)}_\nu(k R \gg 1) \approx \sqrt{2/\pi\,k R}\,e^{i\left(k R - \frac{\pi\nu}{2} - \frac{\pi}{4}\right)}$. For short-range responses, see Appendix~\ref{ap_2}. 
 
 In RKKY–type interactions, the energy integral is dominated by excitations near the Fermi level. To see this, one typically rewrites the energy variable in terms of the momentum (or wave number) $k$. This implies that near the Fermi energy $\mathcal{E}_{\rm F}-\hbar^2\,\Delta\,\Omega_{\rm d}/m\,\lambda^2$, accounting for the light-induced energy–offset, one can rewrite $k_{\pm} = k_{\rm F}/\sqrt{1\pm\beta} \approx k^{\pm}_{\rm F}$. After straightforward calculations, for $k_{\rm F}R \gg 1$, one obtains\begin{widetext}\begin{subequations}\label{eq_17}
       \begin{align}
            \widetilde{\mathcal{J}}_{xx}(R) \approx {} & -\frac{1}{R^2 (1-\beta^2)} \left\{(1+\beta)
\cos(2k^+_{\rm F}R-\tfrac{\pi}{4})
+(1-\beta)
\cos(2k^-_{\rm F}R-\tfrac{\pi}{4})-2\,\beta\cos(2\phi)\sin([k^+_{\rm F}+k^-_{\rm F}]R)\right\}\, ,\label{eq_16a}\\
                \widetilde{\mathcal{J}}_{zz}(R) \approx {} & -\frac{1}{R^2 (1-\beta^2)} \left\{(1+\beta)
\cos(2k^+_{\rm F}R-\tfrac{\pi}{4})
+(1-\beta)
\cos(2k^-_{\rm F}R-\tfrac{\pi}{4})+2\,\beta\cos(2\phi)\sin([k^+_{\rm F}+k^-_{\rm F}]R)\right\}\, ,\label{eq_16b}\\
                \widetilde{\mathcal{J}}_{{\rm DM},x}(R) \approx  {} &+ \frac{\lambda_y\,\sin(\phi)}{R^2 (1-\beta^2)}\left\{
(1+\beta)
\sin(2k^+_{\rm F}R-\tfrac{\pi}{4})
-(1-\beta)
\sin(2k^-_{\rm F}R-\tfrac{\pi}{4})\right\}\,,\label{eq_16c}\\
                \widetilde{\mathcal{J}}_{{\rm DM},y}(R) \approx  {} &+ \frac{\lambda_x\,\cos(\phi)}{R^2 (1-\beta^2)}\left\{
(1+\beta)
\sin(2k^+_{\rm F}R-\tfrac{\pi}{4})
+(1-\beta)
\sin(2k^-_{\rm F}R-\tfrac{\pi}{4})\right\}\,,\label{eq_16d}
            \end{align}
        \end{subequations}\end{widetext}where $\widetilde{\mathcal{J}} = \mathcal{J}/J_l$ with $J_l = \frac{J^2 m^2}{16\pi^2 \hbar^4}$. As previously mentioned, gating is crucial for detecting the effects of light. However, this holds only for anisotropic and antisymmetric DM interactions, as seen in Eqs.~\eqref{eq_16c} and~\eqref{eq_16d} through $\lambda_y$ and $\lambda_x$, respectively. In the limit where \( \beta = 0 \) and \( \Delta = 0 \), i.e., in the absence of light and altermagnetism features, the results of the RKKY interaction in a disordered 2D electron gas and low-dimensional electrons with RSOC are recovered~\cite{PhysRevB.82.165303,10.1063/1.2817405}.\begin{figure}[t]
        \centering
        \includegraphics[width=0.8\linewidth]{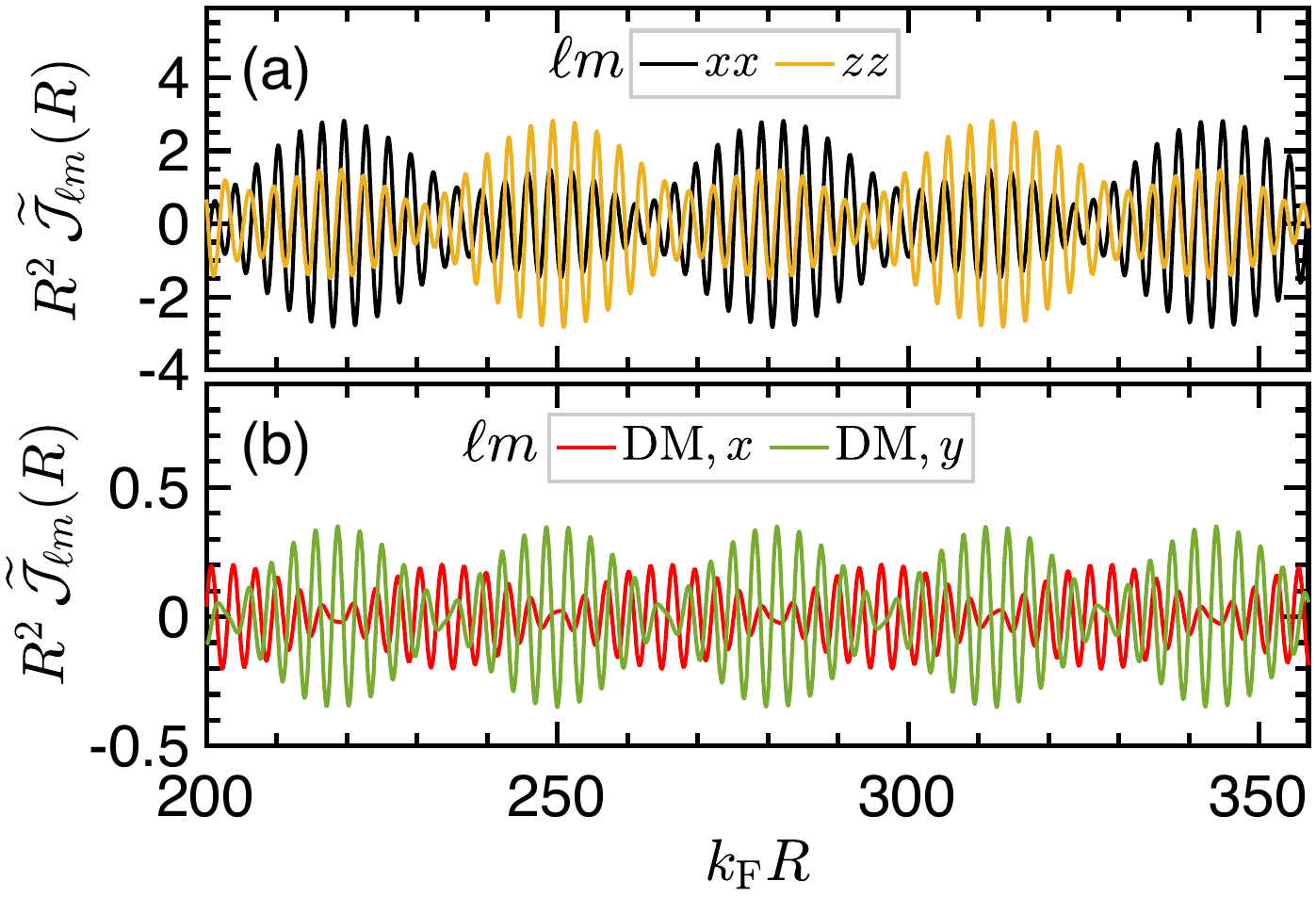}
        \caption{The noncollinear beating-type RKKY interactions in an undriven-gated 2D $d$-wave altermagnet with \(\beta = 0.1\), \(\lambda = 0.2\) eV$\cdot$\AA, and \(\Delta = 0\) eV reveal a striking dominance of the Heisenberg and Ising interactions~[panel (a)], followed by the contributions from the antisymmetric DM interactions~[panel (b)]. } 
        \label{f2}
    \end{figure}

        The direct consequence of these expressions is that all types of RKKY interactions in a 2D $d$-wave altermagnet decay as \( R^{-2} \) when the impurities are far apart, which is expected for 2D systems. Therefore, we focus on the coherent signals, which are strongly influenced by the altermagnetism strength \(\beta\) and the light exchange energy \(\Delta\). They all oscillate with the impurity separation $R$, with periods of \(\pi/k^{\pm}_{\rm F}\), which depend on how the system is doped to modify the Fermi energy and on the specific target altermagnet~(characterized by $\beta$) being considered. 
        
        It is also important to note that these expressions display a unique angular dependence on \( \phi \)~(the direction of the separation vector between the impurities in the $xy$-plane). Specifically, when the magentic moment of one impurity is rotated relative to the other, the interaction oscillates and shows periodic behavior. Therefore, we choose to fix the angle at a specific value, such as \(\phi = \pi/6\), which is arbitrary, with no physical reasoning behind it other than having a non-zero term. We also set $\hbar = m = 1$ in the following data for simplicity. 

    To understand the competition of the interactions in the overall RKKY interaction response, let us start with the undriven-gated 2D $d$-wave altermagnet, $\Delta = 0$ eV, in Fig.~\ref{f2}. We set the bare RSOC to a weak value of \(\lambda = 0.2\) eV$\cdot$\AA\, hereafter to support our analytical expressions. From Eq.~\eqref{eq_17}, beating-type oscillations are expected, which arise as a direct consequence of the intricate interplay between the orientation of magnetic impurities and the propagation of itinerant electrons in the host altermagnet near the Fermi level~\cite{PhysRevB.110.054427}. However, the Heisenberg and Ising interactions dominate the other contributions, with the anisotropic/antisymmetric DM interaction being 10 and 20 times weaker along the \(y\) and \(x\) directions, respectively. \begin{figure}[b]
			\centering
		\includegraphics[width=0.8\linewidth]{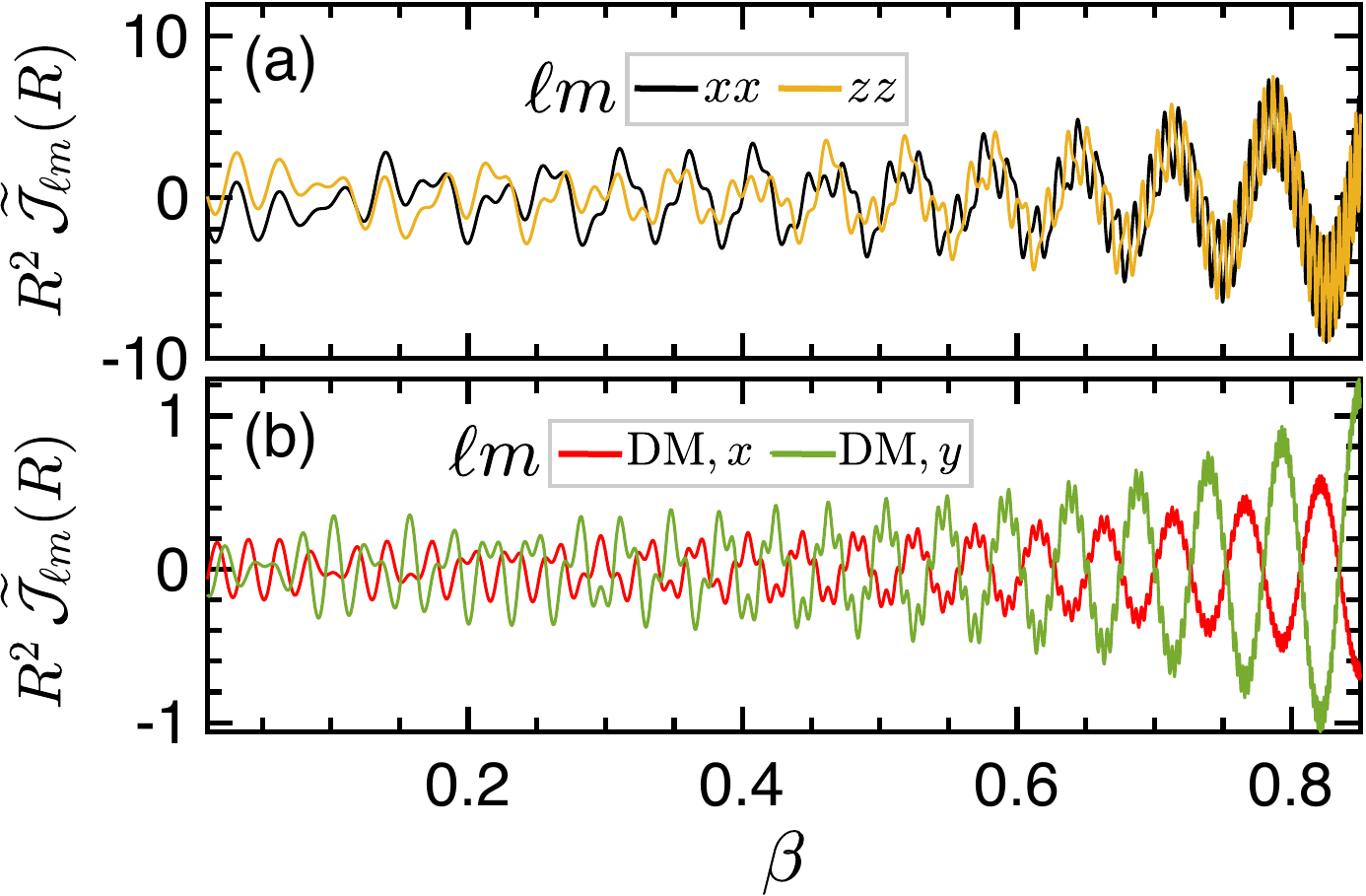}
			\caption{Chirp-like signals~($\beta$ dependency) for the RKKY interactions in an undriven-gated 2D \(d\)-wave altermagnet with \(\lambda=0.2\) eV$\cdot$\AA\, and \(k_{\rm F}R=275\) exhibit isotropic behavior for the Heisenberg and Ising interactions in the strong altermagnetic regime [panel (a)]. In contrast, the anisotropic and antisymmetric DM interaction displays markedly anisotropic amplitude oscillations along the \(x\) and \(y\) directions across all altermagnetic regimes [panel (b)].} 
			\label{f3}
		\end{figure} 

        Next, in Fig.~\ref{f3}, we see how altermagnetism strength, $\beta$, tunes the RKKY interactions. This parameter can be tuned in real materials by doping, applying strain, applying magnetic fields, altering temperature, changing layer thickness, or selecting specific materials with various intrinsic spin-splittings~\cite{PhysRevX.12.040501,PhysRevX.12.031042,Si2016,PhysRevB.109.144421}. These methods offer a range of possibilities for controlling the behavior of altermagnets in practical applications. Strong altermagnetism strength means that the electrons’ spins will be strongly polarized in momentum space even though the material as a whole does not generate an external magnetic field. Altermagnetism strength causes the RKKY signals to exhibit a chirp-like behavior, ``sweeping'' through a range of \(\beta\) values, adapting to the evolving oscillations. For weak altermagnetism, all interactions respond differently. However, for strong altermagnetism, Heisenberg and Ising interactions respond similarly, while we still observe different DM responses. For practical applications, this can be particularly useful in magnetization~(magnetization of the relative orientation between the impurity spins) reversal in spintronics, where the system must overcome an energy barrier. This means the inherent chirp-like signal exhibits a range of oscillatory frequencies across space, influenced by the tunable strength of altermagnetism $\beta$. This variability allows the signal to be adjusted to match the specific frequency needed for magnetization switching, ensuring that energy is delivered precisely at the right moments to help the magnetization surpass the barrier. \begin{figure}[b]
        	\centering
        	\includegraphics[width=0.8\linewidth]{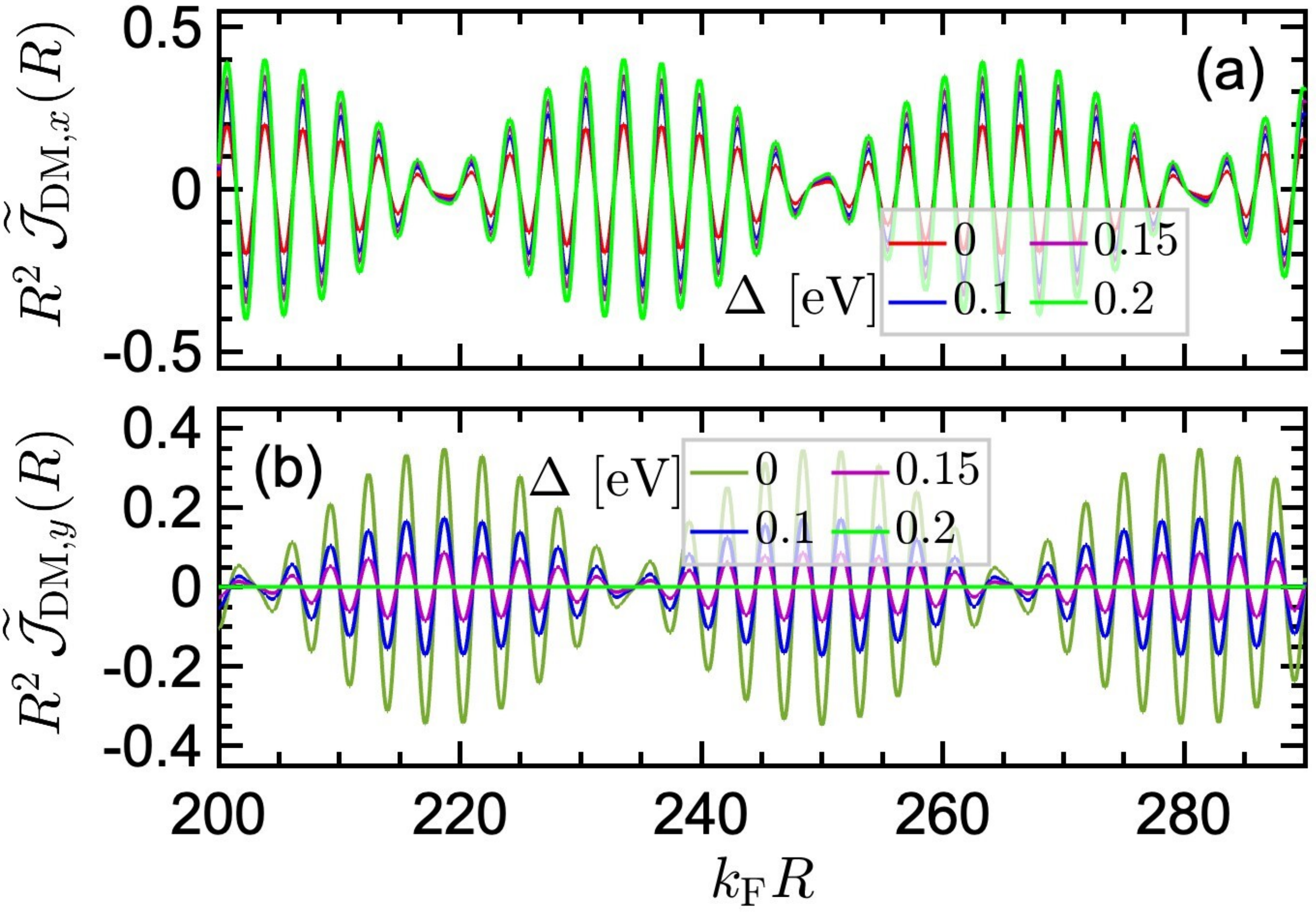}
        	\caption{Light-induced modulation of DM interactions in a driven-gated 2D \(d\)-wave altermagnet with \(\lambda=0.2\) eV$\cdot$\AA\, and \(\beta=0.1\) exhibits anisotropy, with an enhanced \(x\)-component [panel (a)] and a diminished \(y\)-component [panel (b)].} 
        	\label{f6new}
        \end{figure}%The inherent chirp-like signal in altermagnets can be engineered so that its instantaneous frequency aligns with the magnetization’s changing resonant frequency for the switching process as it approaches and crosses the barrier. 

        To examine how light affects the interactions, we analyzed all RKKY interactions for different values of $\Delta$ in Fig.~\ref{f6new}. From Eqs.~\eqref{eq_16a} and~\eqref{eq_16b}, no $\Delta$ dependence is expected for Heisenberg and Ising interactions. This is because the light influences only the gate-induced RSOC, which contributes exclusively to the anisotropic and antisymmetric DM interactions. However, the amplitude of the DM signals increases (decreases) when the light tunes the RSOC along the $x$-direction ($y$-direction). Equations~\eqref{eq_16c} and~\eqref{eq_16d} predict these trends with varying amplitudes and signs in their oscillations, which is confirmed in Figs.~\ref{f6new}(a) and~\ref{f6new}(b), respectively. It should be noted that, for \( \Delta = 0.2 \) eV, the RSOC along the \( x \)-direction, \( \lambda_x = \lambda - \frac{2\Delta \beta}{\lambda} \), vanishes when \( \beta = 0.1 \) and \( \lambda = 0.2 \) eV·\AA. Consequently, the component \( \widetilde{\mathcal{J}}_{{\rm DM},y} \) vanishes at \( \Delta = 0.2 \) eV.%We observe that only DM interactions can be adjusted in a 2D $d$-wave altermagnet, though anisotropically, using light.

        We next turn our attention to see how DM interactions evolve under the influence of light, quantified by \(\Delta\). In Fig.~\ref{f4}(a), the evolution is shown for a fixed altermagnetism strength, \(\beta = 0.4\). Notably, both in-plane components of the DM interaction increase with rising light amplitude. Note that the negative sign of $\widetilde{\mathcal{J}}_{{\rm DM},y}$ should not make any confusions about the anisotropic responses to the light. However, while the \(x\)-component exhibits a steady, monotonic increase without any sign reversal, the \(y\)-component undergoes a sign inversion at a specific threshold. This sign change indicates that the effective coupling between the magnetic impurities reverses its rotational symmetry~(spin chirality). More precisely, this reversal occurs when the \(y\)-component, \(\widetilde{\mathcal{J}}_{{\rm DM},y}\), vanishes, which defines the critical light amplitude as\begin{equation}\label{eq_18}
    \mathcal{E}_0^{\rm c} = {} \frac{2}{e\,\lambda} \,\Omega_{\rm d}^{3/2}\,\Delta^{{\rm c}^{1/2}}\,.
\end{equation}For \(\beta = 0.4\) and $\lambda = 0.2$ eV$\cdot$\AA, this condition is met at \(\Delta^{\rm c} = 0.05\) eV, in excellent agreement with the trends displayed in Fig.~\ref{f4}(a). This results in $\mathcal{E}_0^{\rm c} \approx 4 \times 10^{9}$ V/m, a strength that can now be achieved in experiments~\cite{doi:10.1126/science.1239834,McIver2020,PhysRevResearch.2.043408,Higuchi2017,liu2024evidencefloquetelectronicsteady}. In experiments, the estimated and applied electric field strength depends on various factors such as loss energy. Thus, using 10$^9$ V/m for a free-standing model at zero temperature does not pose the same challenges encountered in experiments~\cite{Higuchi2017,liu2024evidencefloquetelectronicsteady}.\begin{figure}[t]
			\centering
		\includegraphics[width=0.8\linewidth]{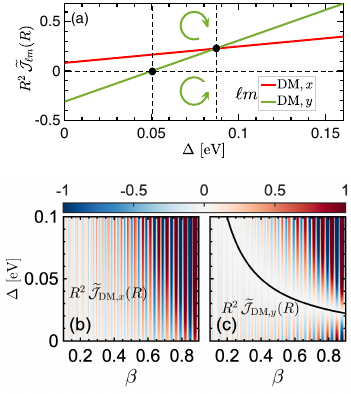}
			\caption{The anisotropic and antisymmetric RKKY interactions in a driven-gated 2D $d$-wave altermagnet with \(\lambda = 0.2\) eV$\cdot$\AA\, and \(k_{\rm F} R = 275\). In panel (a), we set \(\beta = 0.4\). The DM interaction along the \(y\)-axis vanishes at $\Delta = 0.05$ eV, as derived in Eq.~\eqref{eq_18}. On the other hand, the in-plane DM interactions become equal at \(\Delta \approx 0.088\) eV, according to Eq.~\eqref{eq_19}. The phase diagrams in panels (b) and (c) illustrate the anisotropic modulation of the DM components. In (c), we reflect the systematic relationship between light and the altermagnetism along the \(y\)-direction, Eq.~\eqref{eq_18}.} 
			\label{f4}
		\end{figure} 

Furthermore, our analysis indicates that the two components converge to an equivalent value at a distinct critical light strength, $\Delta^{\rm c}_2 \approx 0.088$ eV. This equivalence is derived by carefully considering Eqs.~\eqref{eq_16c} and~\eqref{eq_16d} such that the solution of the following relation marks the point at which the two DM interaction components are equal:{\small\begin{align}\label{eq_19}
    \frac{[1+\cot(\phi)] + \frac{2 \Delta^{\rm c}_2 \beta}{\lambda^2\,\Omega_{\rm d}}[1-\cot(\phi)]}{[1-\cot(\phi)] + \frac{2 \Delta^{\rm c}_2 \beta}{\lambda^2\,\Omega_{\rm d}}[1+\cot(\phi)]} =  \frac{(1-\beta)\sin(2k^-_{\rm F}R-\frac{\pi}{4})}{(1+\beta)\sin(2k^+_{\rm F}R-\frac{\pi}{4})}\, .
\end{align}}

Finally, Figs.~\ref{f4}(b) and~\ref{f4}(c) illustrate a systematic exploration of the DM interactions as both the light exchange energy, \(\Delta\), and the altermagnetism strength, \(\beta\), are varied simultaneously. Our analysis reveals that the amplitude of the \(x\)-component, \(\widetilde{\mathcal{J}}_{{\rm DM},x}\), increases steadily with higher values of both \(\Delta\) and \(\beta\). In contrast, the \(y\)-component, \(\widetilde{\mathcal{J}}_{{\rm DM},y}\), displays a more complex behavior: its amplitude initially decreases as \(\Delta\) increases, up to the given critical threshold. Beyond this threshold, \(\widetilde{\mathcal{J}}_{{\rm DM},y}\) undergoes a sign reversal and then increases in amplitude. This behavior highlights the intricate interplay between light and altermagnetism in tuning the DM interactions anisotropically. We briefly note that all these effects can be reversed~($\widetilde{\mathcal{J}}_{{\rm DM},x/y} \to \widetilde{\mathcal{J}}_{{\rm DM},y/x}$) when using left circularly polarized light. See Appendix~\ref{ap_0} for details.

\section{Summary}\label{s5}
In conclusion, our study unveils a pathway for the dynamic control of RKKY interactions in 2D $d$-wave altermagnets through the application of circularly polarized light. We demonstrate that in these 2D systems, the magnetic landscape of RKKY interaction is predominantly governed by robust Heisenberg and Ising exchange interactions, complemented by a significant yet subtler DM interaction. The inherent altermagnetism strength imparts chirp-like signatures to the magnetic responses. What is particularly intriguing is that these chirp-like responses can be finely adjusted using light. 

This work demonstrates that applying an external gate to induce Rashba spin-orbit coupling is not just a minor tweak—it is essential for controlling the DM interaction under light illumination. The gate-induced Rashba coupling acts in a highly selective way, modifying only the DM interaction anisotropically while leaving the fundamental (primary) exchange interactions intact. This anisotropic control means that the DM interaction, which typically influences the twist or rotation between adjacent impurities, can be tuned differently along various directions in the material when exposed to optical stimuli. Our study also demonstrates that by rotating the altermagnet’s crystallographic orientation by 45$^\circ$ relative to the illuminated surface, light can effectively induce a Dirac-like dispersion and different DM interactions. 

The study further identifies a critical point in the altermagnetic strength. When this threshold is reached, the effect of light on the DM interaction changes—the rotation induced by the DM interaction along the $y$-axis reverses direction. This implies that by carefully tuning the level of altermagnetism (and by extension, the optical conditions), one can control not only the magnitude but also the direction of the spin twisting.

The ability to use light to drive such precise changes in the RKKY interactions provides deep insights into the underlying magnetic phenomena in altermagnets. More importantly, this light-driven control mechanism opens up new possibilities for engineering custom magnetic properties. Such advances are crucial for the development of next-generation spintronic devices, where manipulating electron spins with high precision is key to enhancing performance and functionality.

\section*{Acknowledgments}
    M.Y. greatly thanks Masoumeh Davoudiniya and Peter M. Oppeneer for helpful discussions and acknowledges the hospitality of Uppsala University during his visit, where part of this work was performed. M. Y. and J.K.F. were supported by the Department of Energy, Office of Basic Energy Sciences, Division of Materials Sciences and Engineering under Contract No. DE-FG02-08ER46542 for the formal developments, the numerical work, and the writing of the manuscript. J.L. was supported by the Research Council of Norway through its Centres of Excellence funding scheme Grant No. 262633. J.B. acknowledges support by the DFG through SFB TRR227, B06, and Project No. 465098690.
}
\onecolumngrid
\appendix
{\allowdisplaybreaks
\section{Effective Hamiltonian of a linearly polarized light-induced 2D $d$-wave altermagnet}\label{ap_0l}
In this Appendix, following a linearly polarized vector potential $\vec{A}(t) = A_0\, [\sin(\Omega_{\rm d}\, t),0]$ and minimal coupling regime $\vec{k} \to \vec{k} - e\vec{A}(t)$, we use the van Vleck expansion to obtain the light-induced effective Hamiltonian in 2D $d$-wave altermagnets. We start with \small  \begin{align}
	    \mathcal{H}^{\rm R}_{\vec k- e \vec{A}}(t) = &\alpha_{\vec{k}} \sigma_0 + \beta_{\vec{k}}\, \sigma_z + \lambda\, (k_x\,\sigma_y - k_y\,\sigma_x) + \frac{\hbar^2}{2\,m} \left[e^2\,A^2_0\sin^2(\Omega_{\rm d}t)-2\,e\,A_0\,k_x\sin(\Omega_{\rm d}t)\right][\sigma_0+\beta\,\sigma_z]- e\,A_0\,\lambda\,\sin(\Omega_{\rm d}t) \sigma_y.
            \end{align}One needs to proceed with the following expression to find the light-induced Hamiltonian~\cite{PhysRevB.84.235108,PhysRevB.82.235114,PhysRevB.84.235108,PhysRevLett.110.026603}: $ \mathcal{H}^{\rm eff}_{\vec{k}} \simeq {} \mathcal{H}^{\rm F}_0 +\tfrac{\left[\mathcal{H}^{\rm F}_{-1},\mathcal{H}^{\rm F}_{+1}\right]}{\Omega_{\rm d}}$, where $\mathcal{H}^{\rm F}_0 = {} \frac{1}{T} \int^T_0 \mathcal{H}^{\rm R}_{\vec k- e \vec{A}}(t)\, dt$ and $\mathcal{H}^{\rm F}_{\pm 1} = {} \frac{1}{T} \int^T_0 \mathcal{H}^{\rm R}_{\vec k- e \vec{A}}(t)\, e^{\pm i \,\Omega_{\rm d} t} dt$. For $\mathcal{H}^{\rm F}_0$, we find\begin{align}
    \mathcal{H}^{\rm F}_0 = {} & \left(\alpha_{\vec{k}} + \frac{\hbar^2\,e^2\,A_0^2}{4\,m}\right)\, \sigma_0 + \left(\beta_{\vec{k}} + \frac{\hbar^2\,e^2\,A_0^2\,\beta}{4\,m}\right)\, \sigma_z + \lambda\, (k_x\,\sigma_y - k_y\,\sigma_x)\, .
\end{align}Moreover, for $\mathcal{H}^{\rm F}_{\pm 1}$, we achieve\begin{align}
        \mathcal{H}^{\rm F}_{\pm 1} = {} \mp i\frac{\hbar^2\,e\,A_0}{2\,m}\,k_x[\sigma_0 + \beta\,\sigma_z] \mp i \frac{e\,A_0\,\lambda}{2}\,\sigma_y\, ,
    \end{align}resulting in $[\mathcal{H}^{\rm F}_{-1},\mathcal{H}^{\rm F}_{+1}]/\Omega_{\rm d} = {} 0$ and\begin{align}
    \mathcal{H}^{\rm eff}_{\vec{k}} \simeq {} & \left(\alpha_{\vec{k}} + \frac{\hbar^2\,e^2\,A_0^2}{4\,m}\right)\, \sigma_0 + \left(\beta_{\vec{k}} + \frac{\hbar^2\,e^2\,A_0^2\,\beta}{4\,m}\right)\, \sigma_z + \lambda\, (k_x\,\sigma_y - k_y\,\sigma_x)\, .
\end{align}Thus, a linearly polarized light only shifts the chemical potential and adds a magnetization to the altermagnetism term, which generally results in a rescaling of the system's energies, without affecting the RSOC term, which is important for practical spintronic applications. The same procedure can be applied to \(\vec{A}(t) = A_0 \, [0, \cos(\Omega_{\rm d} t)] \), which leads to the same contributions.

 \section{Effective Hamiltonian of an elliptically polarized light-induced 2D $d$-wave altermagnet}\label{ap_0e}
In this Appendix, following an elliptically polarized vector potential $\vec{A}(t) = [A_x\sin(\Omega_{\rm d}\, t),A_y\cos(\Omega_{\rm d}\, t+\delta)]$, where $A_x$ and $A_y$ are the amplitudes of the $x$ and $y$ components, respectively, while $\delta$ is the phase difference between the two components. We employ the minimal coupling regime $\vec{k} \to \vec{k} - e\vec{A}(t)$ and use the van Vleck expansion to obtain the Floquet terms in the effective Hamiltonian. We start with
\begin{align}
	\mathcal{H}^{\rm R}_{\vec k- e \vec{A}}(t) = {} &\alpha_{\vec{k}}\, \sigma_0 + \beta_{\vec{k}}\, \sigma_z + \lambda\, (k_x\,\sigma_y - k_y\,\sigma_x) + \frac{\hbar^2}{2\,m}\Big[e^2\,A_x^2\,\sin^2(\Omega_{\rm d}\,t)+ e^2\,A_y^2\,\cos^2(\Omega_{\rm d}\,t+\delta)\notag\\
	- {} & 2 e \left(A_x\,k_x\sin(\Omega_{\rm d}t)+A_y\,k_y\cos(\Omega_{\rm d}t+\delta)\right)\Big]\sigma_0\, + \frac{\hbar^2 \beta}{2\,m}\Big[e^2\,A_x^2\,\sin^2(\Omega_{\rm d}\,t)- e^2\,A_y^2\,\cos^2(\Omega_{\rm d}\,t+\delta) \notag\\
	- {} & 2 e \left(A_x\,k_x\sin(\Omega_{\rm d}t)-A_y\,k_y\cos(\Omega_{\rm d}t+\delta)\right)\Big]\sigma_z - \lambda e  \left(A_x\,\sin(\Omega_{\rm d}t) \sigma_y-A_y\,\cos(\Omega_{\rm d}t+\delta)\sigma_x\right)\, .
\end{align}For $\mathcal{H}^{\rm F}_0 = {} \frac{1}{T} \int^T_0 \mathcal{H}^{\rm R}_{\vec k- e \vec{A}}(t)\, dt$, we have\begin{align}
	\mathcal{H}^{\rm F}_0 = {} & \left(\alpha_{\vec{k}}+\frac{\hbar^2\,e^2\,[A_x^2+A_y^2] }{4\,m}\right)\, \sigma_0 +\left( \beta_{\vec{k}}+\frac{\hbar^2\,e^2\,[A_x^2-A_y^2]\,\beta}{4\,m}\right)\, \sigma_z + \lambda\, (k_x\,\sigma_y - k_y\,\sigma_x)\, .
\end{align}Moreover, for $\mathcal{H}^{\rm F}_{\pm 1} = {} \frac{1}{T} \int^T_0 \mathcal{H}^{\rm R}_{\vec k- e \vec{A}}(t)\, e^{\pm i \,\Omega_{\rm d} t} dt$, we achieve\begin{align}
	\mathcal{H}^{\rm F}_{\pm1} = {} -\frac{\hbar^2\,e}{2\,m}(A_y\,k_y\pm i\,A_x\,k_x)\sigma_0 + \frac{\hbar^2\,\beta\,e}{2\,m}(A_y\,k_y\mp i\,A_x\,k_x)\sigma_z + \frac{\lambda\,e}{2}(A_y\,\sigma_x \mp i\,A_x\,\sigma_y)\, ,
\end{align}resulting in
\begin{align}
	\frac{[\mathcal{H}^{\rm F}_{-1},\mathcal{H}^{\rm F}_{+1}]}{\Omega_{\rm d}} = {} \frac{\lambda^2\,e^2\,A_x\,A_y}{\Omega_{\rm d}}\, \sigma_z - \frac{\hbar^2\,e^2\,A_x\,A_y\,\lambda \,\beta}{m\,\Omega_{\rm d}}\, (k_x\,\sigma_y +k_y\,\sigma_x)\, .
\end{align}Substituting the above equations into \(\mathcal{H}^{\rm eff}_{\vec{k}} \simeq \mathcal{H}^{\rm F}_0 + \frac{\left[\mathcal{H}^{\rm F}_{-1}, \mathcal{H}^{\rm F}_{+1}\right]}{\Omega_{\rm d}}\) results in\begin{align}\label{eq_b5}
	\mathcal{H}^{\rm eff}_{\vec{k}} \simeq {} & \left(\alpha_{\vec{k}}+\frac{\hbar^2\,e^2\,[A_x^2+A_y^2] }{4\,m}\right)\, \sigma_0 + \left(\beta_{\vec{k}}+\frac{\hbar^2\,e^2\,[A_x^2-A_y^2]\,\beta}{4\,m}+\frac{\lambda^2\,e^2\,A_x\,A_y}{\Omega_{\rm d}}\right)\, \sigma_z \notag \\ + {} & \left[\left(\lambda- \frac{\hbar^2\,e^2\,A_x\,A_y\,\lambda \,\beta}{m\,\Omega_{\rm d}}\right)\, k_x\,\sigma_y - \left(\lambda+ \frac{\hbar^2\,e^2\,A_x\,A_y\,\lambda \,\beta}{m\,\Omega_{\rm d}}\right)k_y\,\sigma_x \right] .
\end{align}

\section{Effective Hamiltonian of a circularly polarized light-induced 2D $d$-wave altermagnet}\label{ap_0}
In this Appendix, following a right circularly polarized vector potential $\vec{A}(t) = A_0\, [\sin(\Omega_{\rm d}\, t),\cos(\Omega_{\rm d}\, t)]$ and minimal coupling regime $\vec{k} \to \vec{k} - e\vec{A}(t)$, we use the van Vleck expansion to obtain the Floquet terms in the effective Hamiltonian given by Eq.~\eqref{eq_5}. We start with
    \begin{align}
	    \mathcal{H}^{\rm R}_{\vec k- e \vec{A}}(t) = {} &\alpha_{\vec{k}}\, \sigma_0 + \beta_{\vec{k}}\, \sigma_z + \lambda\, (k_x\,\sigma_y - k_y\,\sigma_x) + \frac{\hbar^2}{2\,m}\left[e^2\,A_0^2 - 2 e A_0\left(k_x\sin(\Omega_{\rm d}t)+k_y\cos(\Omega_{\rm d}t)\right)\right]\sigma_0\, \notag\\
        + {} & \frac{\hbar^2 \beta}{2\,m}\left[-e^2\,A_0^2\cos(2\Omega_{\rm d}t) - 2 e A_0\left(k_x\sin(\Omega_{\rm d}t)-k_y\cos(\Omega_{\rm d}t)\right)\right]\sigma_z - \lambda e A_0 \left(\sin(\Omega_{\rm d}t) \sigma_y-\cos(\Omega_{\rm d}t)\sigma_x\right)\, .
            \end{align}For $\mathcal{H}^{\rm F}_0 = {} \frac{1}{T} \int^T_0 \mathcal{H}^{\rm R}_{\vec k- e \vec{A}}(t)\, dt$, we have\begin{align}
    \mathcal{H}^{\rm F}_0 = {} & \left(\alpha_{\vec{k}}+\frac{\hbar^2\,e^2\,A_0^2 }{2\,m}\right)\, \sigma_0 + \beta_{\vec{k}}\, \sigma_z + \lambda\, (k_x\,\sigma_y - k_y\,\sigma_x)\, .
\end{align}Moreover, for $\mathcal{H}^{\rm F}_{\pm 1} = {} \frac{1}{T} \int^T_0 \mathcal{H}^{\rm R}_{\vec k- e \vec{A}}(t)\, e^{\pm i \,\Omega_{\rm d} t} dt$, we achieve\begin{align}
        \mathcal{H}^{\rm F}_{\pm1} = {} -\frac{\hbar^2\,e\,A_0}{2\,m}(k_y\pm ik_x)\sigma_0 + \frac{\hbar^2\,\beta\,e\,A_0}{2\,m}(k_y\mp ik_x)\sigma_z + \frac{\lambda\,e\,A_0}{2}(\sigma_x \mp i\sigma_y)\, ,
    \end{align}resulting in
\begin{align}
    \frac{[\mathcal{H}^{\rm F}_{-1},\mathcal{H}^{\rm F}_{+1}]}{\Omega_{\rm d}} = {} \frac{\lambda^2\,e^2\,A^2_0}{\Omega_{\rm d}}\, \sigma_z - \frac{\hbar^2\,e^2\,A^2_0\,\lambda \,\beta}{m\,\Omega_{\rm d}}\, (k_x\,\sigma_y +k_y\,\sigma_x)\, .
\end{align}Substituting the above equations into \(\mathcal{H}^{\rm eff}_{\vec{k}} \simeq \mathcal{H}^{\rm F}_0 + \frac{\left[\mathcal{H}^{\rm F}_{-1}, \mathcal{H}^{\rm F}_{+1}\right]}{\Omega_{\rm d}}\) and defining $\mathcal{A}_0 = e A_0 \lambda$ and $\Delta = \mathcal{A}_0^2/2\,\Omega_{\rm d}$ results in\begin{align}\label{eq_b5}
   \mathcal{H}^{\rm eff}_{\vec{k}} \simeq {} \left(\alpha_{\vec{k}}+\frac{\hbar^2\,\Delta\,\Omega_{\rm d}}{m\,\lambda^2}\right)\, \sigma_0 + \left(\beta_{\vec{k}}+2\Delta\right)\, \sigma_z + \left[\left(\lambda- \frac{2\hbar^2\,\Delta \,\beta}{m\,\lambda}\right)\, k_x\,\sigma_y - \left(\lambda+ \frac{2\hbar^2\,\Delta \,\beta}{m\,\lambda}\right)k_y\,\sigma_x \right] .
    \end{align}For the left circularly polarized vector potential $\vec{A}(t) = A_0\, [-\sin(\Omega_{\rm d}\, t),\cos(\Omega_{\rm d}\, t)]$, we find a sign change $\Delta \to -\Delta$ in the second and third terms as\begin{align}\label{eq_b5}
    \mathcal{H}^{\rm eff}_{\vec{k}} \simeq {} \left(\alpha_{\vec{k}}+\frac{\hbar^2\,\Delta\,\Omega_{\rm d}}{m\,\lambda^2}\right)\, \sigma_0 + \left(\beta_{\vec{k}}-2\Delta\right)\, \sigma_z + \left[\left(\lambda+ \frac{2\hbar^2\,\Delta \,\beta}{m\,\lambda}\right)\, k_x\,\sigma_y - \left(\lambda- \frac{2\hbar^2\,\Delta \,\beta}{m\,\lambda}\right)k_y\,\sigma_x \right] .
    \end{align}

\section{Derivation of closed-form analytical expressions for $G_n(R,\mathcal{E})$ for $n\in\{0,x,y,z\}$}\label{ap_1nn}
In this Appendix, we present the detailed derivations of the real-space Green's functions for light-induced and gated 2D $d$-wave altermagnets. We consider the relation $G(R, \mathcal{E}) = {} G_0(R, \mathcal{E})\,\sigma_0 + \vec{G}(R, \mathcal{E})\cdot \vec{\sigma}$ with diagonal~($G_0(R, \mathcal{E})$,$G_z(R, \mathcal{E})$) and off-diagonal~($G_x(R, \mathcal{E})$,$G_y(R, \mathcal{E})$) elements.
\subsection{Diagonal $G_0(R,\mathcal{E})$}
First, we solve the following integral:
\begin{align}
G_0(R,\mathcal{E})= {} \frac{1}{4\pi^2}\int d^2 k\,e^{i\vec{k}\cdot \vec{R}}
\,\frac{\mathcal{E}+i\eta-\dfrac{\hbar^2}{2m}(k_x^2+k_y^2)-\frac{\hbar^2\,\Delta\,\Omega_{\rm d}}{m\,\lambda^2}}
{\left[\mathcal{E}+i\eta-\tfrac{\hbar^2}{2m}(k_x^2+k_y^2)-\frac{\hbar^2\,\Delta\,\Omega_{\rm d}}{m\,\lambda^2}\right]^2-\left(\tfrac{\hbar^2\beta}{2m}(k_x^2-k_y^2)\right)^2}\, ,
\end{align}where in our notation, the momentum and real–space vectors are written in polar form as $
\vec{k}=(k\cos\theta,k\sin\theta)$ and $\vec{R}=(R\cos\phi,R\sin\phi)$. A key observation is that the numerator factors nicely with the denominator and leads to a partial fraction expansion. In fact, one may verify that
\begin{equation}
\frac{A(k)}{A(k)^2-B(k)^2\cos^2 2\theta}
=\frac{1}{2}\left[
\frac{1}{A(k)-B(k)\cos2\theta}+
\frac{1}{A(k)+B(k)\cos2\theta}
\right].
\label{eq:partial_fraction}
\end{equation}where $A(k) = \mathcal{E}+i\eta-\frac{\hbar^2k^2}{2m}-\frac{\hbar^2\,\Delta\,\Omega_{\rm d}}{m\,\lambda^2}$ and $B(k) = \frac{\hbar^2\beta\,k^2}{2m}$. In each term the angular integration is of the standard form $\int_0^{2\pi}d\theta\,\frac{e^{ikR\cos\theta}}{A(k)\pm B(k)\cos2\theta}$. One can then expand the exponential as $e^{ikR\cos\theta}=\sum_{\ell=-\infty}^{+\infty} i^\ell J_\ell(kR)e^{i \ell \theta}$, and use the standard result
\begin{align}
\int_0^{2\pi}d\theta\,\frac{e^{i\ell\theta}}{A(k)\pm B(k)\cos2\theta}
=\frac{2\pi}{\sqrt{A(k)^2-B(k)^2}}\left(\frac{A(k)-\sqrt{A(k)^2-B(k)^2}}{\pm B(k)}\right)^{|\ell|},
\end{align}with appropriate interpretation when \(\ell\) is not an even integer. After solving the simplified integrals, $G_0(R,\mathcal{E})$ can be written in terms of Hankel functions as
\begin{equation}
G_0(R,\mathcal{E}) = {} \frac{-m}{4\hbar^2\,\sqrt{1-\beta^2}}
\left[H_0^{(1)}(k_+R)+H_0^{(1)}(k_-R)\right]\, ,
\label{eq:final}
\end{equation}with $k_\pm=\sqrt{\frac{2m\tilde{\mathcal{E}}}{\hbar^2(1\pm\beta)}}$ with $\tilde{\mathcal{E}} = \mathcal{E}-\frac{\hbar^2\,\Delta\,\Omega_{\rm d}}{m\,\lambda^2}$. In writing this result we have taken the limit $\eta\to0^+$. The appearance of the two distinct wave numbers \(k_+\) and \(k_-\) is a consequence of the anisotropic dispersion introduced by the altermagnetism parameter \(\beta\). 

\subsection{Off-diagonal $G_x(R,\mathcal{E})$}

Second, we wish to evaluate the two–dimensional integral
\begin{align}
G_x(R,\mathcal{E}) = {} \frac{1}{4\pi^2}\int d^2 k\,e^{i\vec{k} \cdot \vec{R}}\,\frac{\lambda_y\,k_y}{\Bigl[\mathcal{E}+i\eta-\frac{\hbar^2(k_x^2+k_y^2)}{2m}-\frac{\hbar^2\,\Delta\,\Omega_{\rm d}}{m\,\lambda^2}\Bigr]^2-\Bigl(\frac{\hbar^2\beta\,(k_x^2-k_y^2)}{2m}\Bigr)^2}\,,
\end{align}
where the integration is carried out over the entire (positive) $k$–plane using the polar coordinate representation:
%\[ k_x=k\cos\theta,\quad k_y=k\sin\theta,\quad R_x=R\cos\phi,\quad R_y=R\sin\phi,\quad k=\sqrt{k_x^2+k_y^2}\,.\] Here $\lambda_y$, $\beta$, and $E$ are parameters and $\delta>0$ is an infinitesimal real number.
\begin{align}
G_x(R,\mathcal{E}) = {} \frac{\lambda_y}{4\pi^2}\int_0^\infty k^2\,dk\int_0^{2\pi}d\theta\, \frac{e^{ikR\cos(\theta-\phi)}\,\sin\theta}{A(k)^2-B(k)^2\cos^2 2\theta}\,.
\end{align}Following the same procedure, the expression for $G_x(R,\mathcal{E})$ yields
\begin{align}
G_x(R,\mathcal{E})=\frac{\lambda_y}{4\pi^2}\int_0^\infty k^2\,dk\sum_{\ell=-\infty}^\infty i^\ell J_\ell (kR)e^{-i\ell\phi}\int_0^{2\pi}d\theta\, \frac{e^{i\ell \theta}\,\sin\theta}{A(k)^2-B(k)^2\cos^2 2\theta}\,.
\end{align}
Since the denominator depends on $\theta$ only through $\cos2\theta$, it can be expanded in a Fourier series containing only even harmonics: $[A(k)^2-B(k)^2\cos^2 2\theta]^{-1}=\sum_{\ell'=-\infty}^\infty C_{2\ell'}(k)e^{2i\ell'\theta}$ with coefficients $C_{2\ell'}(k) = \frac{(-1)^{\ell'} }{A(k)}\left(\frac{B(k)}{2 A(k)}\right)^{|2\ell'|} \frac{\Gamma(\ell' + 1/2)}{\Gamma(1/2)\Gamma(\ell' + 1)}$, which $\Gamma(\dots)$ is the Gamma function. Next, the angular integral becomes $\int_0^{2\pi}d\theta\, e^{i(\ell+2\ell')\theta}\,\sin\theta=\frac{1}{2i}\Bigl[2\pi\,\delta_{\ell+2\ell'+1,\,0}-2\pi\,\delta_{\ell+2\ell'-1,\,0}\Bigr]
=\frac{\pi}{i}\Bigl[\delta_{\ell+2\ell',-1}-\delta_{\ell+2\ell',1}\Bigr]$. Thus only those combinations of $\ell$ and $\ell'$ satisfying $\ell+2\ell'=\pm 1$ contribute. One may then re–sum the resulting series (after a shift in the summation index) and show that the net effect is to select the $+1$ cylindrical–wave component. A further (standard) contour integration over $k$ (whose details we omit) yields the final result in terms of Hankel functions of order one:
\begin{equation}
G_x(R,\mathcal{E})= {} \frac{-i\,\lambda_y\, m\,e^{i\phi}}{4\hbar^4\sqrt{1-\beta^2}}\,
\left[\frac{H_1^{(1)}(k_+R)}{k_+}-\frac{H_1^{(1)}(k_-R)}{k_-}\right]\,.
\label{eq:final_Iy}
\end{equation}where the factor \(ie^{i\phi}\) arises from the angular dependence of the numerator \(k_y\propto\sin\theta\).

\subsection{Off-diagonal $G_y(R,\mathcal{E})$}
Third, we wish to evaluate
\begin{align}
G_y(R,\mathcal{E}) = {} \frac{1}{4\pi^2}\int d^2 k\,e^{i\vec{k} \cdot \vec{R}}\,\frac{\lambda_x\,k_x}{\Bigl[\mathcal{E}+i\eta-\frac{\hbar^2(k_x^2+k_y^2)}{2m}-\frac{\hbar^2\,\Delta\,\Omega_{\rm d}}{m\,\lambda^2}\Bigr]^2-\Bigl(\frac{\hbar^2\beta\,(k_x^2-k_y^2)}{2m}\Bigr)^2}\,,
\end{align}leading straightforwardly to\begin{align}
G_y(R,\mathcal{E})=\frac{\lambda_x}{4\pi^2}\int_0^\infty k^2\,dk\sum_{\ell,\ell'} i^\ell J_\ell(kR)e^{-i\ell\phi}\,C_{2\ell'}(k)
\int_0^{2\pi}d\theta\, e^{i(\ell+2\ell')\theta}\,\cos\theta\,.
\end{align}A somewhat lengthy (omitted here, but standard) calculation shows that the final answer may be written in closed form as
\begin{equation}
\label{eq:final}
G_y(R,\mathcal{E})= {} \frac{-\lambda_x\, m\,e^{i\phi}}{4\hbar^4\sqrt{1-\beta^2}}\,
\left[\frac{H_1^{(1)}(k_+R)}{k_+}+\frac{H_1^{(1)}(k_-R)}{k_-}\right]\, .
\end{equation}

\subsection{Diagonal $G_z(R,\mathcal{E})$}

Fourth, we wish to evaluate the 2D integral
\begin{equation}
\label{eq:I}
G_z(R,\mathcal{E}) = \frac{1}{4\pi^2}\int d^2 k\,e^{i \vec{k} \cdot \vec{R}}\,
\frac{\frac{\hbar^2\beta\,(k_x^2-k_y^2)}{2m}}
{\Bigl[\mathcal{E}+i\eta-\frac{\hbar^2(k_x^2+k_y^2)}{2m}-\frac{\hbar^2\,\Delta\,\Omega_{\rm d}}{m\,\lambda^2}\Bigr]^2-\Bigl(\frac{\hbar^2\beta\,(k_x^2-k_y^2)}{2m}\Bigr)^2}\,,
\end{equation}By applying the same procedure as outlined previously, the integral becomes
\begin{equation}
\label{eq:Ipolar}
G_z(R,\mathcal{E})=\frac{1}{8\pi^2 }\int_0^\infty k^3\,dk\int_0^{2\pi}d\theta\,e^{ikR\cos(\theta-\phi)}\,\left[
\frac{1}{A(k)-B(k)\cos2\theta}-
\frac{1}{A(k)+B(k)\cos2\theta}
\right]\,.
\end{equation}Again, using the Fourier--Bessel series and standard contour–integration techniques, one finds that the final answer may be written in closed form in terms of Hankel functions:\begin{equation}
\begin{aligned}
G_z(R,\mathcal{E}) = {} \frac{-m\,\beta\,\cos(2\phi)}{4\hbar^2\sqrt{1-\beta^2}}\,
\left[\frac{H_2^{(1)}(k_+R)}{k_+^2}+\frac{H_2^{(1)}(k_-R)}{k_-^2}\right]\, .
\end{aligned}
\label{eq:final}
\end{equation}

 \section{A circularly polarized light-induced 45$^\circ$ rotated 2D $d$-wave altermagnet}\label{ap_00}
In this Appendix, we repeat the calculations of Appendix~\ref{ap_0} for a 45$^\circ$ rotated 2D $d$-wave altermagnet with the Hamiltonian $\mathcal{H}_{\vec k} = \alpha_{\vec{k}}\, \sigma_0 + \frac{\hbar^2 \beta}{m} k_x\,k_y\,\sigma_z+\lambda\, (k_x\,\sigma_y - k_y\,\sigma_x)$, where the RSOC term is invariant and the altermagnetism term is rotated from \( \beta_{\vec{k}} = \frac{\hbar^2\beta}{2\,m}(k_x^2 - k_y^2) \) to \( \theta_{\vec{k}} = \frac{\hbar^2 \beta}{m} k_x k_y \). We start with
\begin{align}
	&\mathcal{H}^{\rm R}_{\vec k- e \vec{A}}(t) = {} \alpha_{\vec{k}}\, \sigma_0 + \theta_{\vec{k}}\, \sigma_z + \lambda\, (k_x\,\sigma_y - k_y\,\sigma_x) + \frac{\hbar^2}{2\,m}\left[e^2\,A_0^2 - 2 e A_0\left(k_x\sin(\Omega_{\rm d}t)+k_y\cos(\Omega_{\rm d}t)\right)\right]\sigma_0\notag\\
	{} & +\frac{\hbar^2 \beta}{m}\left[\frac{e^2\,A_0^2}{2}\sin(2\,\Omega_{\rm d}\,t)- e A_0\left(k_x\cos(\Omega_{\rm d}t)+k_y\sin(\Omega_{\rm d}t)\right)\right]\sigma_z - \lambda e A_0 \left(\sin(\Omega_{\rm d}t) \sigma_y-\cos(\Omega_{\rm d}t)\sigma_x\right)\, .
\end{align}For $\mathcal{H}^{\rm F}_0 = {} \frac{1}{T} \int^T_0 \mathcal{H}^{\rm R}_{\vec k- e \vec{A}}(t)\, dt$, we have\begin{align}
	\mathcal{H}^{\rm F}_0 = {} & \left(\alpha_{\vec{k}}+\frac{\hbar^2\,e^2\,A_0^2 }{2\,m}\right)\, \sigma_0 + \theta_{\vec{k}}\, \sigma_z +\lambda\, (k_x\,\sigma_y - k_y\,\sigma_x)\, .
\end{align}Moreover, for $\mathcal{H}^{\rm F}_{\pm 1} = {} \frac{1}{T} \int^T_0 \mathcal{H}^{\rm R}_{\vec k- e \vec{A}}(t)\, e^{\pm i \,\Omega_{\rm d} t} dt$, we achieve\begin{align}
	\mathcal{H}^{\rm F}_{\pm 1} = {}  -\frac{\hbar^2\,e\,A_0}{2\,m}(k_y\pm ik_x)\sigma_0 - \frac{\hbar^2\,\beta\,e\,A_0}{m}(k_x\pm ik_y)\sigma_z + \frac{\lambda\,e\,A_0}{2}(\sigma_x \mp i\sigma_y)\,,
\end{align}resulting in
\begin{align}
	\frac{[\mathcal{H}^{\rm F}_{-1},\mathcal{H}^{\rm F}_{+1}]}{\Omega_{\rm d}} = {} \frac{\lambda^2\,e^2\,A^2_0}{2\,\Omega_{\rm d}}\, \sigma_z - \frac{2\,\hbar^2\,e^2\,A^2_0\,\lambda \,\beta}{m\,\Omega_{\rm d}}\, (k_x\,\sigma_x +k_y\,\sigma_y)\, .
\end{align}Substituting the above equations into \(\mathcal{H}^{\rm eff}_{\vec{k}} \simeq \mathcal{H}^{\rm F}_0 + \frac{\left[\mathcal{H}^{\rm F}_{-1}, \mathcal{H}^{\rm F}_{+1}\right]}{\Omega_{\rm d}}\) and defining $\mathcal{A}_0 = e A_0 \lambda$ and $\Delta = \mathcal{A}_0^2/2\,\Omega_{\rm d}$ results in\begin{align}
	\mathcal{H}^{\rm eff}_{\vec{k}} \simeq {} &\left(\alpha_{\vec{k}}+\frac{\hbar^2\,\Delta\,\Omega_{\rm d}}{m\,\lambda^2}\right)\, \sigma_0 + \left(\theta_{\vec{k}}+\Delta\right)\, \sigma_z +\left[\left(\lambda\,k_x- \frac{4\,\hbar^2\,\Delta \,\beta}{m\,\lambda}\,k_y\right)\,\sigma_y - \left(\lambda\,k_y+ \frac{4\,\hbar^2\,\Delta \,\beta}{m\,\lambda}\,k_x\right)\,\sigma_x\right]\,.
\end{align}The light-induced magnetization is halved due to the 45\(^\circ\) rotation. Additionally, such a 45$^\circ$ rotation induces Dirac-like spectrum $\vec{k} \cdot \vec{\sigma}$ in a 2D $d$-wave altermagnet.\begin{figure}[t]
	\centering
	\includegraphics[width=1\linewidth]{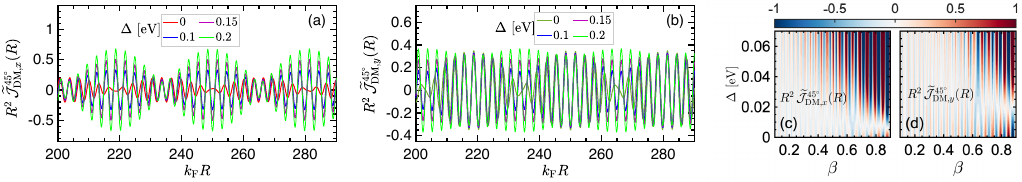}
	\caption{[(a),(b)] The evolution of a driven‐gated 45$^\circ$‐rotated 2D \(d\)-wave altermagnet (\(\lambda=0.2\) eV\(\cdot\)\AA, \(\beta=0.1\)) with light. This mechanism modulates the DM interactions, yielding enhanced components compared to the 0$^\circ$ configuration. [(c),(d)] Systematic variation of light exchange energy and altermagnetism strength produces multiple rotational reversals in both components, in contrast to the uniaxial changes observed at 0$^\circ$ rotation.} 
	\label{f7n}
\end{figure} 

In this case, retarded real-space Green's functions \( G_0(R,\mathcal{E}) \) and \( G_z(R,\mathcal{E}) \) remain unchanged, while the other two off-diagonal components are modified as
\begin{subequations}
	\begin{align}
		\widetilde{G}_x(R,\mathcal{E})= &\frac{-me^{i\phi}}{4\hbar^4\sqrt{1-\beta^2}}
		\left[\left(i\,\lambda+\frac{4\hbar^2\,\Delta\,\beta}{\lambda}\right)\frac{H_1^{(1)}(k_+R)}{k_+}-\left(i\,\lambda-\frac{4\hbar^2\,\Delta\,\beta}{\lambda}\right)\frac{H_1^{(1)}(k_-R)}{k_-}\right]\,,\\
		\widetilde{G}_y(R,\mathcal{E})=  &\frac{-me^{i\phi}}{4\hbar^4\sqrt{1-\beta^2}}
		\left[\left(\lambda-i\,\frac{4\,\hbar^2\,\Delta\,\beta}{\lambda}\right)\frac{H_1^{(1)}(k_+R)}{k_+}+\left(\lambda+i\frac{4\hbar^2\,\Delta\,\beta}{\lambda}\right)\frac{H_1^{(1)}(k_-R)}{k_-}\right]\, .
	\end{align}
\end{subequations}Accordingly, the interactions \( \widetilde{\mathcal{J}}_{xx}(R) \) and \( \widetilde{\mathcal{J}}_{zz}(R) \) remain unchanged, while for the DM interactions, we find\begin{subequations}\label{eq_20n}
	\begin{align}
		\widetilde{\mathcal{J}}^{45^\circ}_{{\rm DM},x}(R) \approx  {} &+ \frac{1}{R^2 (1-\beta^2)}\Bigg\{\tilde{\lambda}_1(1+\beta)
		\sin(2k^+_{\rm F}R-\tfrac{\pi}{4})-\tilde{\lambda}_2(1-\beta)
		\sin(2k^-_{\rm F}R-\tfrac{\pi}{4})\Bigg\}\,,\label{eq_20na}\\
		\widetilde{\mathcal{J}}^{45^\circ}_{{\rm DM},y}(R) \approx  {} &+ \frac{1}{R^2 (1-\beta^2)}\Bigg\{\tilde{\lambda}_3(1+\beta)
		\sin(2k^+_{\rm F}R-\tfrac{\pi}{4})+\tilde{\lambda}_4(1-\beta)
		\sin(2k^-_{\rm F}R-\tfrac{\pi}{4})\Bigg\}\,,\label{eq_20nb}
	\end{align}
\end{subequations}where $\tilde{\lambda}_1 = \left[-
\lambda\,\sin(\phi)+\tfrac{4\,\hbar^2\,\Delta\,\beta}{\lambda}\cos(\phi)\right]$, $\tilde{\lambda}_2 = \left[-
\lambda\,\sin(\phi)-\tfrac{4\,\hbar^2\,\Delta\,\beta}{\lambda}\cos(\phi)\right]$, $ \tilde{\lambda}_3 = \left[
\lambda\,\cos(\phi)-\tfrac{4\,\hbar^2\,\Delta\,\beta}{\lambda}\sin(\phi)\right]$, and $\tilde{\lambda}_4 = \left[
\lambda\,\cos(\phi)+\tfrac{4\,\hbar^2\,\Delta\,\beta}{\lambda}\sin(\phi)\right]$.

In driven‐gated 45$^\circ$‐rotated 2D \(d\)-wave altermagnets (with \(\lambda=0.2\) eV\(\cdot\)\AA\ and \(\beta=0.1\)) the application of light induces a modulation of the DM interactions that yields enhanced components relative to the 0$^\circ$ configuration, as demonstrated in Figs.~\ref{f7n}(a) and~\ref{f7n}(b). Moreover, systematic variation of the light energy and the strength of altermagnetism in Figs.~\ref{f7n}(c) and~\ref{f7n}(d) results in multiple rotational reversals in both components, in marked contrast to the uniaxial modifications observed at 0$^\circ$ rotation. This complex behavior underscores the sensitivity of the external light field to the crystallographic orientation of the altermagent, offering new avenues for tuning spin textures in altermagnetic materials.

 \section{Short-range RKKY interactions}\label{ap_2}
In this Appendix, we present the short-range RKKY interactions using the following series expansions:
\begin{align}
    H^{(1)}_0(k R \ll 1) \approx {}  1 + \frac{2i}{\pi} \ln(kR/2)\qquad , \qquad
    H^{(1)}_1(k R \ll 1) \approx {} \frac{k R}{2} - \frac{2i}{\pi k R}\qquad , \qquad
    H^{(1)}_2(k R \ll 1) \approx {} \frac{k^2 R^2}{8} - \frac{4i}{\pi k^2 R^2} + \frac{1}{2}\, .
\end{align}These expressions should first be applied in Eqs.~\eqref{eq_14} to obtain the short-range real-space Green’s functions. Then, we substitute them into Eqs.~\eqref{eq_13} to determine the RKKY exchange interactions, which involve an energy integration. We stress that the energy integration of the Hankel function itself, from $-\infty$ to 0, is finite, meaning that the full expression, without approximation, does not require regularization, even if it can not be done analytically. However, with the above approximated asymptotic forms, we use high–energy $D$ and a low–energy $\mathcal{E}_{\rm c}$ in numeric for the energy integration. Finally, we find
\begin{subequations}
    \begin{align}
        \widetilde{\mathcal{J}}_{xx}(R) \approx \widetilde{\mathcal{J}}_{zz}(R) = {} & \frac{m D}{\hbar^2(1-\beta^2)}\left\{1-\ln\left(\frac{m D R^2}{2\hbar^2\sqrt{1-\beta^2} }\right)\right\}\, ,\\
        \widetilde{\mathcal{J}}_{{\rm DM},x/y}(R) = {} & -\frac{\beta \lambda_{y/x} \sin(\phi)}{R (1-\beta^2)} \ln(D/\mathcal{E}_{\rm c})\, ,
    \end{align}
\end{subequations}where $\widetilde{\mathcal{J}} = \mathcal{J}/J_s$ with $J_s = \frac{J^2 m}{2\pi^2 \hbar^2}$. Both the Heisenberg and Ising interactions exhibit a similar logarithmic trend with respect to the impurity separation \( R \), while the DM interactions decay with a rate of \( R^{-1} \), as illustrated in Fig.~\ref{f6}(a) and~\ref{f6}(b). However, the evolution with the altermagnetism strength \( \beta \) is governed by a factor of \( 1/(1 - \beta^2) \) for all interactions, as shown in the same panels with blue axes and dotted lines. We also check the evolution of DM interactions with the light energy \( \Delta \) in Fig.~\ref{f6}(c). Similar to the long-range responses, the DM interactions exhibit anisotropic modulation, with \( \widetilde{\mathcal{J}}_{{\rm DM},y} \) vanishing at \(  \Delta^{\rm c} = \lambda^2/2\,\beta \). However, in contrast to the long-range responses (as seen in Fig.~\ref{f4}(a) in the main text), the DM components do not intersect at short-range responses.\begin{figure}[t]
\centering
\includegraphics[width=1\linewidth]{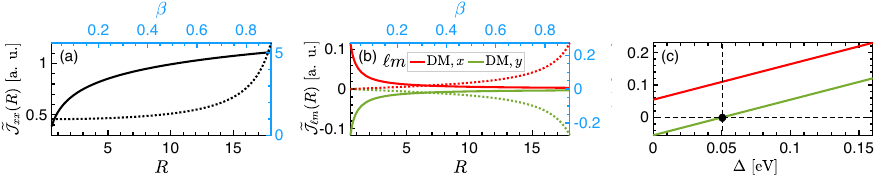}
\caption{Short-range RKKY interactions in a gated 2D $d$-wave altermagnet with $\lambda = 0.1$ eV$\cdot$\AA. The dotted lines show the evolution with respect to \(\beta\) (blue axes), while in panels (a) and (b), we set $\beta = 0.1$ to display the RKKY interaction evolution with respect to the impurity separation $R$ (solid lines). Panel (c) shows the variation of short-range DM interactions with light $\Delta$, where $\beta = 0.4$ and $\lambda = 0.2$ eV$\cdot$\AA\, are used for this analysis.} 
\label{f6}
\end{figure}
}
\twocolumngrid
\bibliography{bib.bib}

\end{document}